\documentclass[review,authoryear,12pt]{elsarticle}

\usepackage{amssymb}

\usepackage{natbib}
\usepackage{amsmath,amssymb,amsfonts}
\usepackage{algorithm2e}
\usepackage{graphicx}
\usepackage{textcomp}
\usepackage{svg}
\usepackage{xcolor}
\usepackage{wrapfig}
\usepackage{subcaption}
\usepackage{multirow}
\usepackage{tabularx,booktabs}
\usepackage{titlesec}
\usepackage{longtable}
\usepackage{lipsum}
\usepackage{listings}
\usepackage[normalem]{ulem}
\usepackage[listings,skins]{tcolorbox}
\usepackage{bm}

\newcommand\norm[1]{\left\lVert#1\right\rVert}

\newcommand{\sgn}{\operatorname{sgn}}
\usepackage{tikz}
\usetikzlibrary{shapes,arrows,positioning,calc,fit}

\DeclareMathOperator*{\argmax}{arg\,max}
\DeclareMathOperator*{\argmin}{arg\,min}

\newcommand\hl[1]{%
  \bgroup
  \hskip0pt\color{red!80!black}%
  #1%
  \egroup
}

 \usepackage{lineno}
 
 \usepackage{multirow}

\journal{Ultrasound in Medicine and Biology}

\begin{document}

\begin{frontmatter}



\title{Ultrasound Signal Processing: \\ From Models to Deep Learning}





\author[Affil1]{Ben~Luijten \corref{cor1}}
\author[Affil1]{Nishith~Chennakeshava \corref{cor2}}
\author[Affil2]{Yonina C. Eldar} 
\author[Affil1]{\\ Massimo Mischi}
\author[Affil1, Affil3]{Ruud J.G. van Sloun}
\address[Affil1]{Dept. of Electrical Engineering, Eindhoven University of Technology, Eindhoven, The Netherlands}
\address[Affil2]{Faculty of Math and CS,
Weizmann Institute of Science, Rehovot, Israel}
\address[Affil3]{Philips Research, Eindhoven}
\cortext[cor1]{Corresponding Author: Ben Luijten, 5612AZ, Eindhoven; Email, w.m.b.luijten@tue.nl; Phone, +31 40 247 9111}
\cortext[cor2]{Corresponding Author: Nishith Chennakeshava, 5612AZ, Eindhoven; Email, n.chennakeshava@tue.nl; Phone, +31 40 247 9111}

\begin{abstract}
    Medical ultrasound imaging relies heavily on high-quality signal processing to provide reliable and interpretable image reconstructions.
    Conventionally, reconstruction algorithms where derived from physical principles. These algorithms rely on assumptions and approximations of the underlying measurement model, limiting image quality in settings were these assumptions break down. Conversely, more sophisticated solutions based on statistical modelling, careful parameter tuning, or through increased model complexity, can be sensitive to different environments.
    Recently, deep learning based methods, which are optimized in a data-driven fashion, have gained popularity. These model-agnostic techniques often rely on generic model structures, and require vast training data to converge to a robust solution. 
    A relatively new paradigm combines the power of the two: leveraging data-driven deep learning, as well as exploiting domain knowledge. These model-based solutions yield high robustness, and require less parameters and training data than conventional neural networks.
    In this work we provide an overview of these techniques from recent literature, and discuss a wide variety of ultrasound applications. We aim to inspire the reader to further research in this area, and to address the opportunities within the field of ultrasound signal processing. We conclude with a future perspective on model-based deep learning techniques for medical ultrasound.

\end{abstract}


\end{frontmatter}

\pagebreak








\section*{Introduction}
\label{sec:intro}
\noindent Ultrasound (US) imaging has proven itself to be an invaluable tool in medical diagnostics. Among many imaging technologies, such as X-ray, computed tomography (CT), and magnetic resonance imaging (MRI), US uniquely positions itself as an interactive diagnostic tool, providing real-time spatial and temporal information to the clinician. Combined with its relatively low cost, compact size, and absence of ionizing radiation, US imaging is an increasingly popular choice in patient monitoring. 

Consequently, the versatility of US imaging has spurred a wide range of applications in the field. While conventionally it is used for the acquisition of B-mode (2D) images, more recent developments have enabled ultrafast, and 3D volumetric imaging. Additionally, US devices can be used for measuring clinically relevant features such as: blood velocity (Doppler), tissue characteristics (e.g. Elastography maps), and perfusion trough ultrasound localization microscopy (ULM). While this wide range of applications shares the same underlying measurement steps: acquisition, reconstruction, and visualisation, their signal processing pipelines are often specific for each application.

It follows that the quality of US imaging strongly depends on the implemented signal processing algorithms. The resulting demand for high-quality signal processing has pushed the reconstruction process from fixed, often hardware based implementations to the digital domain \citep{Thomenius1996, Yongmin1997}. More recently, this has led to fully software-based algorithms, as they can open up the potential to complex measurement models and statistical signal interpretations.
However, this shift has also posed a new set of challenges, as it puts a significant strain on the digitisation hardware, bandwidth constrained data channels, and computational capacity. As a result, clinical devices, where real-time imaging and robustness are of utmost importance, still mainly rely on simple hardware based solutions.

A more recent development in this field is the utilisation of deep neural networks. Such networks can provide fast approximations for signal recovery, and can often be efficiently implemented due to their exploitation of parallel processing. After training, these networks can be efficiently implemented to facilitate ultra-fast signal processing. However, by inheriting generic network architectures from computer vision tasks, these approaches are highly data-driven and are often over-parameterized, posing several challenges. In order to converge to a well-generalised solution across the full data distribution encountered in practice, large amounts of (unbiased) training data are needed, which is not always trivial to obtain. Furthermore, these models are often treated as a `black-box', making it difficult to guarantee the correct behavior in a real clinical setting.

To overcome some of the challenges of purely data-driven methods, an alternative approach is to try to combine model-based and data-driven methods, in attempt to get the best of both worlds. The proposition here is that the design of data-driven methods for ultrasound signal processing can likely benefit from the vast amounts of research on conventional, \textit{model-based}, reconstruction algorithms, informing e.g. specific neural network designs or hybrid processing approaches. 

\begin{figure}
    \centering
    \includegraphics[trim=0 30 0 0, width=1.05\columnwidth]{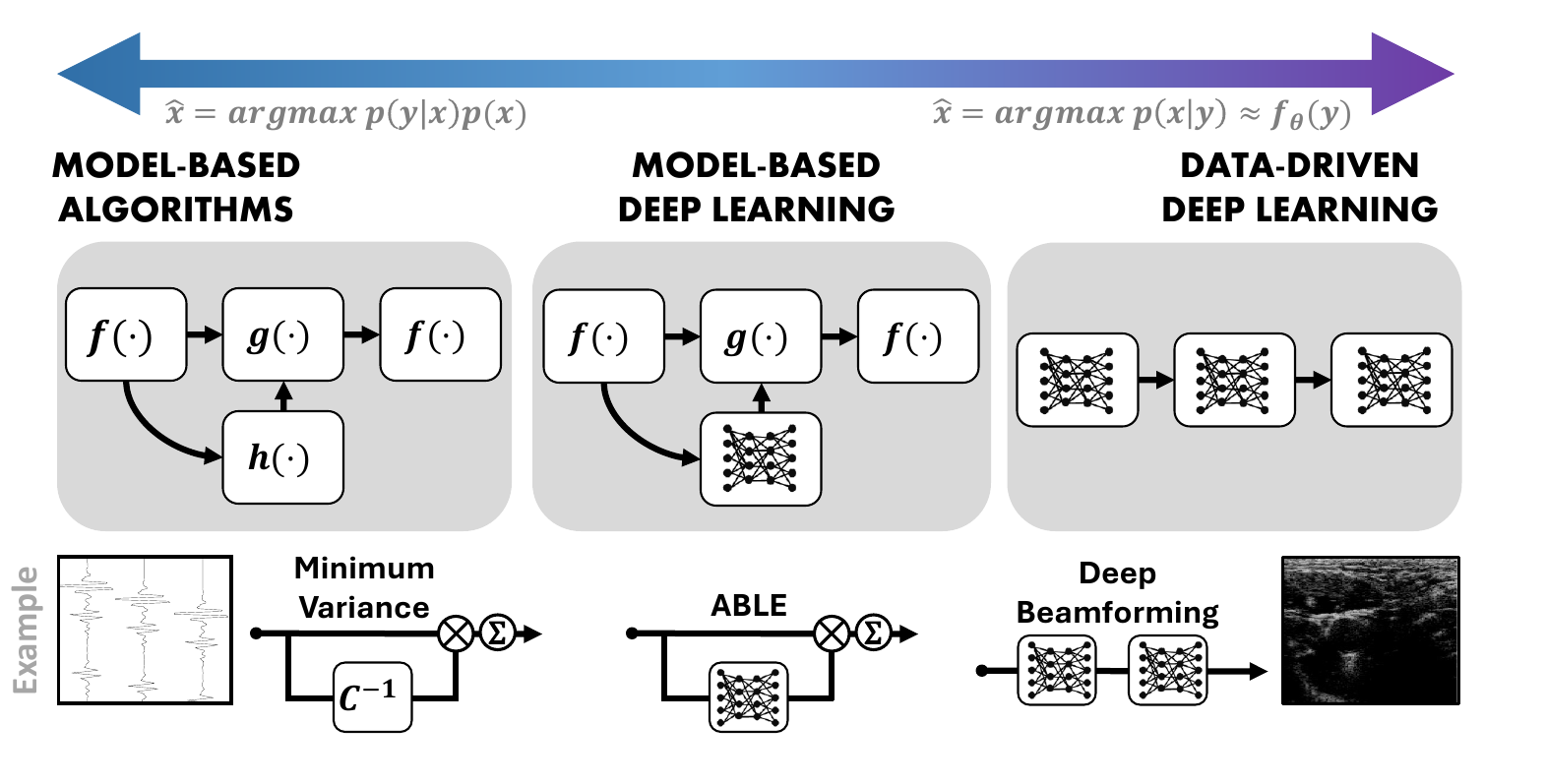}
    \caption{Schematic overview of model-based and deep learning based techniques in ultrasound signal processing. An example for each category is given in the context of ultrasound beamforming: Minimum Variance beamforming, Adaptive Beamforming by Deep Learning (ABLE) \citep{luijten2020adaptive}, and Deep Beamforming \citep{khan2020adaptive}.}
    \label{fig:enter-label}
\end{figure}

In this review paper, we aim to provide the reader with a comprehensive overview of ultrasound signal processing based on modelling, machine learning, and model-based learning. To achieve this, we take a probabilistic perspective and place methods in the context of their assumptions on signal models and statistics, and training data. While other works \citep{shlezinger2020model, Monga2020, van2019deep, al2022review, liu2019deep} offer an excellent overview of the different aspects of AI applied to ultrasound image processing, the focus of this paper is to put the theory of both signal processing and machine learning under a unifying umbrella, rather than to showcase a general review of deep learning being applied to ultrasound specific problems. To that end, we cover topics ranging from beamforming to post-processing and advanced applications such as super-resolution.
Throughout the paper we will distinguish between three types of approaches that we cover in separate sections.
\begin{itemize}
    \item \textbf{Model-Based Methods for US Signal Processing}: Conventional model-based methods derive signal processing algorithms by modelling the problem based on first principles, such as knowledge of the acquisition model, noise, or signal statistics. Simple models offer analytical solutions, while more complex models often require iterative algorithms.
    \item \textbf{Deep Learning (DL) for US Signal Processing}: Deep learning (DL) solutions are fully data-driven and fit highly-parameterized algorithms (in the form of deep neural networks) to data. DL methods are model-agnostic and thus rely on the training data to expose structure and relations between inputs and desired outputs. 
    \item \textbf{Model-Based DL for US Signal Processing}: Model-based DL aims at bridging the gap by deriving algorithms from first-principle models (and their assumptions) while learning parts of these models (or their analytic/iterative solutions) from data. These approaches enable incorporating prior knowledge and structure (inductive biases), and offer tools for designing deep neural networks with architectures that are tailored to a specific problem and setting. The resulting methods resemble conventional model-based methods, but allow for overcoming mismatched or incomplete model information by learning from data.
\end{itemize}

In all cases, data is needed to test the performance of (clinical) signal processing algorithms. However, in deep learning based solutions specifically, we observe an increasing need for training data when prior knowledge on the underlying signal model is not fully exploited.
A schematic overview of these approaches is given in Figure 1, including examples of corresponding techniques in the case of ultrasound beamforming.

We begin by briefly explaining the probabilistic perspective and notation we will adopt throughout the paper in a preliminaries section, after which we provide background information on the basics of US acquisition, which can be skipped by experts in the field of ultrasound. Following this background information, we will dive into model-based US signal processing, in which we will derive various conventional beamforming and post-processing algorithms from their models and statistical assumptions. Next, we turn to DL methods, after which we bridge the gap between model-based and DL-based processing, identifying opportunities for data-driven enhancement of model-based methods (and their assumptions) by DL. Finally we provide a discussion and conclusion, where we provide a future outlook and several opportunities for deep learning in ultrasound signal processing.

\section*{A probabilistic approach to deep learning in ultrasound signal processing}
\label{sec:probabilistic}
\noindent In this paper we will use the language and tools of probability theory to seamlessly bridge the gap between conventional model-based signal processing and contemporary machine/deep learning approaches. As Shakir Mohamed (DeepMind) phrased it:
\begin{center}
\textit{"Almost all of machine learning can be viewed in probabilistic terms, making probabilistic thinking fundamental. It is, of course, not the only view. But it is through this view that we can connect what we do in machine learning to every other computational science, whether that be in stochastic optimisation, control theory, operations research, econometrics, information theory, statistical physics or bio-statistics. For this reason alone, mastery of probabilistic thinking is essential."}
\end{center}
To that end, we begin by briefly reviewing some concepts in probabilistic signal processing based on models, and then turn to recasting such problems as data-driven learning problems.

\subsection*{Preliminaries on model-based probabilistic inference}
\label{sec:prelim_modelbased}

\noindent Let us consider a general linear model
\begin{equation}
    \mathbf{y} = \mathbf{A} \mathbf{x} + \mathbf{n},
    \label{eqn:measurement_model}
\end{equation}
where $\mathbf{y}$ is our observed signal, $\mathbf{A}$ a measurement matrix, $\mathbf{n}$ a noise vector, and $\mathbf{x}$ the signal of interest. As we shall see throughout the paper, many problems in ultrasound signal processing can be described according to such linear models. In ultrasound beamforming for example, $\mathbf{y}$ may denote the measured (noisy) RF signals, $\mathbf{x}$ the spatial tissue reflectivity, and $\textbf{A}$ a matrix that transforms such a reflectivity map to channel domain signals. The goal of beamforming is then to infer $\mathbf{x}$ from $\mathbf{y}$, under the measurement model in \eqref{eqn:measurement_model}.

Recalling Bayes rule, we can define the posterior probability of $\mathbf{x}$ given $\mathbf{y}$, as a product of the likelihood $p(\mathbf{y}|\mathbf{x})$ and a prior $p(\mathbf{x})$, such that
\begin{align}
    p(\mathbf{x}|\mathbf{y}) &= \frac{p(\mathbf{y}|\mathbf{x})p(\mathbf{x})}{p(\mathbf{y})} \\
    &\propto p(\mathbf{y}|\mathbf{x})p(\mathbf{x}).
    \label{eqn:bayes}
\end{align}
Following \eqref{eqn:bayes} we can define a maximum a posteriori (MAP) estimator for \eqref{eqn:measurement_model}, given by
\begin{equation}
    \hat{\mathbf{x}}_\text{MAP} := \argmax_\mathbf{x} p(\mathbf{x}|\mathbf{y}) = \argmax_\mathbf{x} p(\mathbf{y}|\mathbf{x}) p(\mathbf{x}),
    \label{eqn:MAP}
\end{equation}
which provides a single, most likely, estimate according to the posterior distribution. If we assume a Gaussian white noise vector $\mathbf{n}$ in \eqref{eqn:measurement_model}, i.e. $\mathbf{y}\sim \mathcal{N}(\mathbf{A}\mathbf{x}|\bm{\sigma}^2_n I)$, the MAP estimator becomes: 
\begin{equation}
    \hat{\mathbf{x}} = \argmin_\mathbf{x}\norm{\mathbf{y}-\mathbf{A}\mathbf{x}}_{2}^{2} - \lambda\log p(\mathbf{x}),
    \label{eqn:formal_problem}
\end{equation}
where $\lambda$ is a scalar regularization parameter.

Evidently, the MAP estimator takes the prior density function $p(\mathbf{x})$ into account. In other words, it allows us to incorporate and exploit prior information on $\mathbf{x}$, should this be available. 
Conversely, if $x$ is assumed to be deterministic but unknown, we get the maximum likelihood (ML) estimator. The ML estimator thus assigns equal likelihood to each $\mathbf{x}$ in the absence of measurements. As such this simplifies to:
\begin{equation}
    \hat{\mathbf{x}}_\text{ML} := \argmax_x p(\mathbf{y}|\mathbf{x}).
    \label{eqn:ML}
\end{equation}
Many traditional ultrasound processing methods are in this form, where its output only depends on a set of (finely tuned) hyper-parameters, and the input data. This is not surprising, as deriving a strong and useful prior that generalizes well to the entire expected data distribution is challenging in its own right.

Data-driven approaches aim to overcome the challenges of accurate modeling by learning the likelihood function, the prior, the entire posterior, or a direct end-to-end mapping (replacing the complete MAP estimator) from data. We will detail on these methods in the following section.

\subsection*{Preliminaries on deep-learning-based inference}
\label{sec:prelim_DL}

Fully data-driven methods aim at learning the optimal parameters $\theta^*$ of a generic parameterized mapping, $f_\theta (\cdot)$, $Y \rightarrow X$ from training data. 
In deep learning the mapping function $f_\theta (\cdot)$ is a deep neural network. 
Learning itself can also be formulated as a probabilistic inference problem, where optimized parameter settings for a fixed network architecture are inferred from a dataset $\mathcal{D}$. To that end we define a posterior over the parameters:
\begin{align}
p(\theta|\mathcal{D}) &=  \frac{p(\mathcal{D}|\theta)p(\theta)}{p(\mathcal{D})} \\
 & \propto p(\mathcal{D}|\theta)p(\theta),
\end{align}
where $p(\theta)$ denotes a prior over the parameters. Often $p(\theta)$ is fully factorized, i.e. each parameter is assumed independent, to keep the learning problem in deep networks (with millions of parameters) tractable. Typical priors are Gaussian or Laplacian density functions. 

Most deep learning applications rely on MAP estimation to find the set of parameters that minimize the negative log posterior:
\begin{align}
\theta^* &= \argmax_\theta p(\theta|\mathcal{D}) = \argmax_\theta \log p(\mathcal{D}|\theta)p(\theta) \\
  &= \argmin_\theta -\left\lbrace \log p(\mathcal{D}|\theta) + \log p(\theta)\right\rbrace.
\end{align}
Note that for measurement (input) - signal (output) training pairs $(\mathbf{y}_i,\mathbf{x}_i) \in \mathcal{D}$ common forms of $p(\mathbf{x}|f_\theta(\mathbf{y}),\theta)$ are Gaussian, Laplacian, or categorical distributions, resulting in mean-squared-error, mean-absolute-error, and cross-entropy negative log-likelihood functions, respectively. Similarly, Gaussian and Laplacian priors lead to $\ell_2$ and $\ell_1$ regularization on the parameters, respectively. It is worth noting that while most deep learning applications perform MAP estimation, there is increasing interest in so-called Bayesian deep learning, which aims at learning the parameters of the prior distribution $p(\theta)$ as well. This enables posterior sampling during inference (by sampling from $p(\theta)$) for (epistemic) uncertainty estimation. Again, often these distributions are fully factorized (e.g. independent Gaussian or Bernoulli) to make the problem tractable \citep{gal2016dropout}. 

After training (i.e. inferring parameter settings), we can use the network to perform MAP inference to retrieve $\mathbf{x}$ from new input measurements $\mathbf{y}$:
\begin{equation}
    \hat{\mathbf{x}} = \argmax_{\mathbf{x}} p(\mathbf{x}| f_\theta(\mathbf{y}), \theta).
\end{equation}
The neural network thus directly models the parameters of the posterior, and does not factorize it into a likelihood and prior term as model-based MAP inference does. Note that for Gaussian and Laplace density functions, in which a neural network $f_\theta(\mathbf{y})$ computes the distribution mean, $\argmax_{\mathbf{x}} p(\mathbf{x}| f_\theta(\mathbf{y}), \theta) = f_\theta(\mathbf{y})$. For categorical distributions,  $f_\theta(\mathbf{y})$ computes the probabilities for each category/class.

Typical deep neural network parameterizations $f_\theta(\cdot)$ are therefore model-agnostic, as they disregard the structure of the measurement/likelihood model and prior, and offer a high degree of flexibility to fit many data distributions and problems. However, many such parameterizations do exploit specific symmetries in the expected input data. Examples are convolutional neural networks, which exploit the spatial shift-invariant structure of many image classification/regression problems through shift-equivariant convolutional layers. Similarly, many applications where the input is temporally correlated, such as time series analysis, recurrent neural networks (RNN) are employed.

\subsection*{Preliminaries on model-based deep learning}
\label{sec:prelim_modelbasedDL}

\noindent Model-based DL aims at imposing much more structure to the network architectures and parameterizations of $f_\theta(\cdot)$. Where standard deep networks aim at fitting a broad class of problems, model-based DL offers architectures that are highly tailored to specific inference problems given in \eqref{eqn:measurement_model} and \eqref{eqn:MAP} - i.e. they are aware of the model and structure of the problem. This promises to relax challenges related to generalization, robustness, and interpretability in deep learning. It often also enables designing smaller (but more specialized) networks with a lower computational and memory footprint. 

To derive a model-based DL method, one can start by deriving a MAP estimator for $\mathbf{x}$ from the model, including assumptions on likelihood models $p(\mathbf{y}|\mathbf{x})$ and priors $p(\mathbf{x})$. Generally, such estimators come in two forms: \textit{analytic} (direct) and \textit{iterative} solutions. The solution structure dictates the neural network architecture. One then has to select which parts of the original model-based graph are to be replaced by learnable functions. 

One of the first examples of model-based deep learning is the learned iterative-shrinkage and thresholding algorithm (LISTA), proposed by \citet{gregor2010learning}. As the name suggests, $f_\theta(\cdot)$ is based on an iterative solution, specifically to the MAP sparse coding problem: $\argmax_\mathbf{x} p(\mathbf{y}|\mathbf{x})p(\mathbf{x})$, with $\mathbf{x} \sim \text{Laplace}(0,b\mathbf{I})$, where $b$ is a scale parameter, and $\mathbf{y}|\mathbf{x}\sim \mathcal{N}(\mathbf{A} \mathbf{x},\sigma^2 \mathbf{I})$. This iterative solution consists of two alternating steps: 1) a gradient step on $\mathbf{x}$ to maximize the log-likelihood of $\log p(\mathbf{y}|\mathbf{x})$, 2) a proximal step that moves the intermediate solution for $\mathbf{x}$ towards higher log-likelihood under the prior distribution $\log p(\mathbf{x})$. The model-based DL method LISTA \textit{unfolds} or \textit{unrolls} a limited number of algorithm iterations to form a feed-forward neural network, learning the parameters of the gradient step and the proximal step end-to-end from training examples ($\mathbf{y}_i,\mathbf{x}_i \in \mathcal{D}$), without knowledge on the underlying distribution of these parameters. Moreover, LISTA has been shown to accelerate inference compared to the iterative algorithm.

Today, model-based DL is a rapidly developing field, with many variations of such model-based DL methods being developed for various problems and applications \citep{Monga2020}. One can learn optimal gradient steps, adopt neural-network-based proximal operators, and include deep generative models for iterative/unfolded inference, thereby overcoming the limitations of naive assumptions in model-based methods. Similarly, for analytical solutions one can replace computations that rely on accurate model knowledge (that may be unavailable) or are challenging/time-consuming to compute by neural networks. 

Also within the field of US imaging and signal processing, model-based DL is seeing increasing adoption for problems spanning from beamforming \citep{luijten2020adaptive} to clutter suppression \citep{solomon2019deep} and localization microscopy \citep{van2019deep}. Exact implementations of these model-based DL methods for US imaging is indeed highly application specific (which is its merit), as we will discuss in a later section.

\section*{Fundamentals of US acquisition}\label{sec:fundamentals}

\noindent Ultrasound imaging is based on the pulse-echo principle. First, a pressure pulse is transmitted towards a region of interest by the US transducer consisting of multiple transducer elements. Within the medium, scattering occurs due to inhomogeneities in density, speed-of-sound and non-linear behavior. The resulting back-scattered echoes are recorded using the same transducer, yielding a set of radio-frequency (RF) channel signals that can be processed. Typical ultrasound signal processing includes B-mode image reconstruction via beamforming, velocity estimation (Doppler), and additional downstream post-processing and analysis. 

Although the focus of this paper lies on these processing methods, which we will discuss in later chapters, we will for the sake of completeness briefly review the basic principles of ultrasound channel signal acquisition.

\subsection*{Transmit schemes}
\label{sec:transmit}

\noindent Consider an ultrasound transducer with channels $c \in C$. A transmit scheme consists of a series of transmit events $e \in E$. Different transmit events can be constructed by adjusting the per channel transmit delays (focusing), the number of active channels (aperture), and in advanced modes, also waveform parameters. We briefly list the most common transmit schemes. 

\subsubsection*{Line scanning}
\label{sec:linescan}

\noindent Most commercial ultrasound devices rely on focused, line-by-line, acquisition schemes, as it yields superior resolution and contrast compared to unfocused strategies. In line scanning, a subaperture of channels focuses the acoustic energy by channel-dependent transmit delays along a single (axial) path at a set depth, maximizing the reflected echo intensity in a region-of-interest \citep{ding2014ultrasound}. Some transmit schemes make use of multiple foci per line. 
To cover the full lateral field of view, many scan lines are needed, limiting the overall frame rate. 

\subsubsection*{Synthetic aperture}
\label{sec:synthetic_aperture}

\noindent In synthetic aperture (SA) imaging, each channel transmit-receive pair is acquired separately \citep{ylitalo1994ultrasound, jensen2006synthetic}. To that end, each element independently fires a spherical wavefront, of which the reflections can be simultaneously recorded by all receiving elements. Typically, the number of transmit events is equal to the number of transducer elements ($E=C$). Having access to these individual transmit-receive pairs enables retrospective transmit focusing to an arbitrary set of foci (e.g each pixel).
While SA imaging offers advantages in terms of receive processing, it is time consuming, similar to line scanning. Furthermore, single elements generate low acoustic energy, which reduces the SNR.

\subsubsection*{Plane- and Diverging wave}
\label{sec:planewave_divwave}

\noindent As of recent, unfocused (parallel) acquisition schemes have become more popular, since they can drastically reduce acquisition times, yielding so-called ultrafast imaging at very high frame rates.
Plane wave (PW) imaging insonifies the entire region of interest at once through a planar wave field, by firing with all elements and placing the axial focus point at infinity. 
Diverging wave (DW) transmissions also insonify the entire region of interest in one shot, but generate a spherical (diverging) wavefront by placing a (virtual) focus point behind the transducer array. Especially for small transducer footprints (e.g. phased array probes), DW schemes are useful to cover a large image region.

Both PW and DW imaging suffer from deteriorated resolution and low contrast (high clutter) due to strong interference by scattering from all directions. Often, multiple transmits at different angles are therefore compounded to boost image quality. However, this reduces frame rate. Unfocused transmissions rely heavily on the powerful receive processing to yield an image of sufficient quality, raising computational requirements.

\subsubsection*{Doppler}
\label{sec:doppler}

\noindent Beyond positional information, ultrasound also permits measurement of velocities, useful in the context of e.g. blood flow imaging or tissue motion estimation. This imaging mode, called Doppler imaging \citep{chan2011basics, routh1996doppler, hamelmann2019doppler}, often requires dedicated transmit schemes with multiple high-rate sequential acquisitions. 
Continuous wave Doppler allows for simultaneous transmit and receive of acoustic waves, using separate sub-apertures. While this yields a high temporal sampling rate, and prevents aliasing, it does result in some spatial ambiguity. The entire region of overlap between the transmit and receive beam contribute to the velocity estimate.
Alternatively, pulsed-wave Doppler relies on a series of snapshots of the slow-time signal, with the temporal sampling rate being equal to the frame rate. From these measurements, a more confined region-of-interest can be selected for improved position information, at the cost of possible aliasing. 

\subsection*{Waveform and frequency}
\label{sec:waveform}

\noindent The resolution that can be obtained using ultrasound is for a large part dependent on the frequency of the transmitted pulse. High transmit pulse frequencies, and short pulse durations, yield high spatial resolution, but are strongly affected by attenuation. This becomes especially problematic in deep tissue regions. As a general rule, the smallest measurable structures scale to approximately half the wavelength of the transmit frequency, i.e. the diffraction limit. In practice the transmit pulse spans multiple wavelengths, which additionally limits axial resolution by half the transmit pulse length. Design choices such as transducer array aperture, element sensitivity, bandwidth of the front-end circuitry, and reconstruction algorithms also play a dominant role in this.

\subsection*{Array designs}
\label{sec:array}

\noindent Depending on the application, different transducer types may be preferred. Either due to physical constraints, or by having desirable imaging properties. Commonly used transducer geometries include linear-, convex- and phased arrays. 
Effectively, the transducer array, consisting of elements, spatially samples the array response. Typically, these array elements have a center-to-center spacing (pitch) of $\lambda/2$ or less, in order to avoid spatial aliasing.
In general, a higher number of elements yields a better resolution image, but this consequently increases size, complexity, and  bandwidth requirements.
Especially for 2D arrays (used in 3D imaging), the high number of transducer elements can be problematic in implantation due to the vast number of physical hardware connections. Other than translating to an increase in cost and complexity, it also raises power consumption. In those cases, often some form of micro-beamforming is applied in the front-end, combining individual channel signals early in the signal chain. 

Similar reductions in data rates can be achieved through sub-sampling of the receive channels. Trivial approaches include uniform or random sub-sampling, at the cost of reduced resolution, and more pronounced aliasing artifacts (grating lobes).
Several works have showed that these effects can be mitigated either by principled array designs \citep{Cohen2020} \cite{Song2020}, or by learning sub-sampling patterns from data in a task-adaptive fashion \citep{huijben2020learning}.

\subsection*{Sub-Nyquist signal sampling}
\label{sec:compressed_acquisition}

\noindent Digital signal processing of US signals requires sampling of the signals received by the transducer, after which the digital signal is transferred to the processing unit. To prevent frequency-aliasing artifacts, sampling at the Nyquist limit is necessary. In practice, sampling rates of 4-10 times higher are common, as it allows for a finer resolution during digital focusing. As a consequence, this leads to high bandwidth data-streams, which become especially problematic for large transducer arrays (e.g. 3D probes).

Compressed sensing (CS) provides a framework that allows for reduced data rates, by sampling below the Nyquist limit, alleviating the burden on data transfer \citep{Eldar2015sampling}. 
CS acquisition methods provide strong signal recovery guarantees when complemented with advanced processing methods for reconstruction of the signal of interest. These reconstruction methods are typically based on MAP estimation, combining likelihood models on the measured data (i.e. a measurement matrix), with priors on signal structure (e.g. sparsity in some basis). Many of the signal processing algorithms that we will list throughout the paper will find application within a CS context, especially those methods that introduce a signal prior for reconstruction, either through models or by learning from data. The latter is especially useful for elaborate tasks where little is known about the distribution of system parameters, offering signal reconstruction beyond what is possible using conventional CS methods.

For further reading into the fundamentals of ultrasound, the reader may refer to works such as \citet{brahme2014comprehensive}.

\section*{Model-based US signal processing}
\label{sec:model-based}

\noindent Model-based ultrasound signal processing techniques are based on first principles such as the underlying physics of the imaging setup or knowledge of the statistical structure of the signals. We will now describe some of the most commonly used model-based ultrasound signal processing techniques, building upon the probabilistic perspective sketched in earlier sections. For each algorithm we will explicitly list 1) inputs and outputs (and dimensions), 2) the assumed signal model and statistics, 3) signal priors, and 4) the resulting ML/MAP objective and solution.

Beamforming, the act of reconstructing an image from the received raw RF channel signals, is central to ultrasound imaging and typically the first step in the signal processing pipeline. We will thus start our description with beamforming methods.

\subsection*{Beamforming}
\label{sec:model_based_beamforming}

\noindent Given an ultrasound acquisition of $C$ transducer channels, $N_t$ axial samples, and $E$ transmission events, we can denote $Y \in \mathbb{R}^{E \times C \times N_t }$ as the recorded RF data cube, representing back-scattered echoes from each transmission event.
With beamforming, we aim to transform the raw aperture domain signals $Y$ to the spatial domain, through a processing function $f(\cdot)$ such that
\begin{equation}
    \hat{X} = f(Y),
\end{equation}
where $\hat{X}$ represents the data beamformed to a set of focus points $S_\mathbf{r}$. As an example, in pixel-based beamforming, these focus points could be a pixel grid such that $S_\mathbf{r} \in \mathbb{R}^{r_x \times r_z}$, where $r_x$ and $r_y$ represent the lateral and axial components of the vector indicating the pixel coordinates, respectively. Note that, while this example is given in cartesian coordinates, beamforming to other coordinate systems (e.g. polar coordinates) is also common. 

\subsubsection*{Delay-and-sum beamforming}
\label{sec:DAS}

Delay-and-sum (DAS) beamforming has been the backbone of ultrasound image reconstruction for decades. This is mainly driven by its low computational complexity, which allows for real-time processing, and efficient hardware implementations. In DAS, to reconstruct a tissue reflectivity image (B-mode), the aperture domain signals are first migrated back to the image domain, in a process called Time-of-Flight (TOF) correction. This transformation is based on the back-projection of the time-domain signals, and aims at aligning the received signals for a set of focus points (in pixel-based beamforming: pixels) by applying time-delays.
We can define the total TOF from transmission to the receiving element, as
\begin{equation}
    \bm{\tau}_\mathbf{r} = \bm{\tau}[r_x, r_z] = \frac{||\mathbf{r}_{e}-\mathbf{r}||_2+||\mathbf{r}_{c}-\mathbf{r}||_2}{v},
\label{eqn:tof_delays}
\end{equation}
where $\bm{\tau}_\mathbf{r}$ is the required channel delay to focus to an imaging point $\mathbf{r}$, vectors $\mathbf{r}_{e}$ and $\mathbf{r}_{c}$ correspond to the origin of the transmit event $e$, and the position of element $c$, respectively, and $v$ is the speed of sound in the medium. 
Note that the speed-of-sound is generally assumed to be constant throughout the medium. As a consequence, speed-of-sound variations can cause misalignment of the channel signals, and result in aberration errors. 

After TOF correction, we obtain a channel vector $\mathbf{y}_\mathbf{r}$ per pixel $\mathbf{r}$, for which we can define a linear forward model to recover the pixel reflectivity $x_\mathbf{r}$:
\begin{equation}
    \mathbf{y}_\mathbf{r} = \mathbf{1}_\mathbf{r} x_\mathbf{r} + \mathbf{n}_\mathbf{r}
    \label{eqn:narrowband}
\end{equation}
where $\mathbf{y_r} \in \mathbb{R}^C$ is a vector containing the received aperture signals, $x_\mathbf{r} \in \mathbb{R}$ the tissue reflectively at a single focus point $\mathbf{r}$, and $\mathbf{n}_\mathbf{r} \in \mathbb{R}^{C \times 1}$ an additive Gaussian noise vector $\sim \mathcal{N}(0,\sigma_n^2 \mathbf{I})$. In this simplified model, all interference (e.g. clutter, off-axis scattering, thermal noise) is contained in $\mathbf{n}$. Note that (without loss of generality) we assume a real-valued array response in our analysis, which can be straightforwardly extended to complex values (e.g. after in-phase and quadrature demodulation). 
Under the Gaussian noise model, \eqref{eqn:narrowband} yields the following likelihood model for the channel vector:
\begin{equation}
    \begin{aligned}
        p(\mathbf{y_\mathbf{r}}|x_\mathbf{r})
        &=\frac{1}{\sqrt{(2\pi\sigma^2_n)}} \exp \left[ -\frac{(\mathbf{y}_\mathbf{r}-\mathbf{1}_\mathbf{r} x_\mathbf{r})^2}{2\sigma^2_n} \right],
        \label{eqn:das_likelihood}
    \end{aligned}
\end{equation}
where $\sigma^2_n$ denotes the noise power. 

The delay-and-sum beamformer is the per-pixel ML estimator of the tissue reflectively, $\hat{x}_\mathbf{r}$, given by
\begin{align}
        \hat{x}_{\mathbf{r}} :=& \argmax_{x_{\mathbf{r}}} \log  p(\mathbf{y_r}|x_\mathbf{r}) \\ =&  \argmax_{x_{\mathbf{r}}} (\mathbf{y}_\mathbf{r}-\mathbf{1}_\mathbf{r} x_\mathbf{r})^H(\mathbf{y}_\mathbf{r}-\mathbf{1}_\mathbf{r} x_\mathbf{r}).
    \label{eqn:das_optimization}
\end{align}
Solving \eqref{eqn:das_optimization} yields:
\begin{equation}
\hat{x}_\mathbf{r} = \frac{1}{C} \mathbf{1}^H \mathbf{y}_\mathbf{r} = \frac{1}{C}\sum^C_{c=1} y_c,
\label{eqn:ML_DAS}
\end{equation}
where $C$ is the number of array elements. In practice, apodization/tapering weights are included to suppress sidelobes:
\begin{equation}
\hat{x}_\mathbf{r} = \frac{1}{C} \mathbf{w}^H \mathbf{y}_\mathbf{r} = \frac{1}{C}\sum^C_{c=1} w_c y_c.
\label{eqn:ML_DAS_apo}
\end{equation}
This form can be recognized as the standard definition of DAS beamforming, in which the channel signals are weighed using an apodization function, $\mathbf{w}$, and subsequently summed to yield a beamformed signal. 

To promote either higher resolution, or contrast, further adjustments can be made to the apodization function. However, this always poses an inherent compromise between main lobe width and sidelobe intensity. Or equivalently, resolution and contrast. This can be attributed to the beamformer, which is a spatial filter characterized by the aperture pitch (sampling rate), aperture size (filter length), and apodization weights (filter coefficients). Typical choices for apodization weights are tapered functions, such as Hanning or Hamming windows, in which elements further away from the focus point are weighed less then the elements close by. In commercial devices, these apodization weights are often finely tuned to optimize image quality for each transducer type and imaging setting, and may vary spatially (per focus point). 
Similar to DAS for aggregating channels, different transmit events can be aggregated, known as coherent compounding.

\subsubsection*{Frequency domain beamforming}
\label{sec:frequency_beamforming}

In the previous section, all processing was done in the time-domain. Alternatively, the TOF correction can be implemented in the frequency domain.
This has several benefits. Firstly, it avoids the need for oversampling channel data at high-rates to enable accurate time-domain delays. Second, it facilitates sub-Nyquist acquisition schemes that only sample the most important Fourier coefficients (e.g. via sum of sinc filters).
In this context, \citet{Chernyakova2014} propose a Fourier-domain beamformer (FDBF), which allows for a substantial reduction in sampling rates, while maintaining image quality.
Denote by $\omega[k]$ the $k^{th}$ Fourier series coefficient of a beamformed signal. The Fourier transform of  the time-aligned signals, defined as
\begin{equation}
    \hat{\omega}_c [k] = \sum \omega_c [k-n] Q_{k,c,\mathbf{r}}.
\end{equation}
Here $Q_{k,c,\mathbf{r}}$ are the Fourier coefficients of a distortion function derived from the beamforming delays at $\mathbf{r}$, as in \eqref{eqn:tof_delays}. 

When not all Fourier coefficients are sampled (i.e. in sub-Nyquist acquisition), the desired time-domain signal can be recovered using CS methods such as NESTA \cite{Becker2011}, or via deep learning approaches.

\subsection*{Advanced adaptive Beamforming}
\label{sec:advanced_beamforming}

The shortcomings of standard DAS beamforming have spurred the development of a wide range of adaptive beamforming algorithms. These methods aim to overcome some of the limitations that DAS faces, by adaptively tuning its processing based on the input signal statistics.

\subsubsection*{Minimum Variance}
\label{sec:minimum_variance}

DAS beamforming is the ML solution of \eqref{eqn:narrowband} under white Gaussian noise. To improve realism for more structured noise sources, such as off-axis interference, we can introduce a colored (correlated) Gaussian noise profile $\mathbf{n} \sim N(0,\mathbf{\Gamma}_n)$, with $\mathbf{\Gamma}_\mathbf{r}$ being the array covariance matrix for beamforming point $\mathbf{r}$. Maximum (log) likelihood estimation for $x_\mathbf{r}$ then yields:
\begin{align}
    \hat{x}_\mathbf{r} &= \argmax_{x_\mathbf{r}} \log p(\mathbf{y}_\mathbf{r}| x_\mathbf{r},\mathbf{\Gamma}_\mathbf{r}) \\
    &= \argmin_{x_\mathbf{r}} (\mathbf{y}_\mathbf{r}-\mathbf{1}x_\mathbf{r})^H\mathbf{\Gamma}_\mathbf{r}^{-1}(\mathbf{y}_\mathbf{r}-\mathbf{1}x_\mathbf{r}). \label{eqn:MV_ll}
\end{align}
Setting the gradient of the argument in \eqref{eqn:MV_ll} with respect to $\hat{x}_\mathbf{r}$ equal to zero, gives:
\begin{align}
     0 &= \frac{d}{d\hat{x}_\mathbf{r}}(\mathbf{y}_\mathbf{r}-\mathbf{1}\hat{x}_\mathbf{r})^H\mathbf{\Gamma}_\mathbf{r}^{-1}(\mathbf{y}_\mathbf{r}-\mathbf{1}\hat{x}_\mathbf{r}) \\
    0 &=-2\mathbf{1}^H\mathbf{\Gamma}_\mathbf{r}^{-1}(\mathbf{y}_\mathbf{r}-\mathbf{1}\hat{x}_\mathbf{r})\\
   \hat{x}_\mathbf{r} &= ( \mathbf{1}^H\mathbf{\Gamma}_\mathbf{r}^{-1}\mathbf{1})^{-1} \mathbf{1}^H\mathbf{\Gamma}_\mathbf{r}^{-1}\mathbf{y}_\mathbf{r}. \label{eqn:MV_mlestimator}
\end{align}

It can be shown that solution \eqref{eqn:MV_mlestimator} can also be obtained by minimizing the total output power (or variance), while maintaining unity gain in a desired direction (the foresight):

\begin{equation}
\begin{aligned}
\min_\mathbf{w}\mathbf{w}_{\mathbf{r}}^\textrm{H}\mathbf{\Gamma}_\mathbf{r}\mathbf{w}_{\mathbf{r}},\\
s.t.\quad \mathbf{w}_{\mathbf{r}}^\textrm{H}\mathbf{1} = 1.
\end{aligned}
\label{eqn:MV}
\end{equation}
Solving for \eqref{eqn:MV} yields the closed form solution

\begin{equation}
    \hat{\mathbf{w}}_\text{MV} = \frac{\mathbf{1} \Gamma^{-1}_n}{\mathbf{1}^H \Gamma^{-1}_n \mathbf{1}},
    \label{eqn:mv_closedform}
\end{equation}
which is known as Minimum Variance (MV) or Capon beamforming.

In practice, the noise covariance is unknown, and is instead empirically estimated from data ($\mathbf{\Gamma}_n = E[\mathbf{yy}^H]$). For stability of covariance matrix inversion, this estimation often relies on averaging multiple sub-apertures and focus points, or by adding a constant factor to the diagonal of the covariance matrix (diagonal loading). Note here, that for $\mathbf{\Gamma} = \sigma^2_n \mathbf{I}$ (White Gaussian noise), we get the DAS solution as in \eqref{eqn:ML_DAS}.

Minimum Variance beamforming was shown to improve both resolution and contrast in ultrasound images, and has similarly found application in plane wave compounding \citep{Austeng2011}. However, it is computationally complex due to the inversion of the covariance matrix \citep{Raz2002}, leading to significantly longer reconstruction times compared to DAS. To boost image quality further, eigen-space based MV beamforming has been proposed \citep{Deylami2016}, at the expense of further increasing computational complexity.
As a result of this, real-time implementations remain challenging, to an extent that MV beamforming is almost exclusively used as a research tool.

\subsubsection*{Wiener beamforming}
\label{sec:wiener_beamforming}

In the previously covered methods, we have considered the ML estimate of $\hat{x}$. Following \eqref{eqn:MAP}, we can extend this by including a prior probability distribution $p(x_\mathbf{r})$, such that
\begin{equation}
    \hat{x}_\mathbf{r} = \argmax_{x_\mathbf{r}} (\mathbf{y_r}|x_\mathbf{r}) p(x_\mathbf{r}).
    \label{eqn:MAP_BF}
\end{equation}
For a Gaussian likelihood model, the solution to this MAP estimate is equivalent to minimizing the mean-squared-error, such that
\begin{equation}
    \hat{\mathbf{w}} = \argmin_\mathbf{w} E[|x_\mathbf{r}-\mathbf{w}^H\mathbf{y_\mathbf{r}}|^2],
    \label{eqn:mmse}
\end{equation}
also known as Wiener beamforming \citep{vanTrees2004}. Solving this yields
\begin{equation}
\hat{\mathbf{w}}_\text{wiener} = \frac{\sigma^2_x}{\sigma^2_x + \mathbf{w}_\text{MV}^H\Gamma_\mathbf{r}\mathbf{w}_\text{MV}}\mathbf{w}_\text{MV},
\label{eqn:wiener_solution}
\end{equation}
with $\mathbf{\Gamma}_\mathbf{r}$ being the array covariance for beamforming point $\mathbf{r}$, and $\mathbf{w}_\text{MV}$ the MV beamforming weights given by \eqref{eqn:mv_closedform}.
Wiener beamforming is therefore equivalent to MV beamforming, followed by a scaling factor based on the ratio between the signal power and total power of the output signal, which can be referred to as post-filtering.
Based on this result, \citet{Nilsen2010} observe that for any $\mathbf{w}$ that satisfies $\mathbf{w}^H\mathbf{1} = 1$ (unity gain), we can find a Wiener post-filter that minimizes the MSE of the estimated signal. As such, we can write
\begin{align}
    H_\text{wiener} &= \argmin_H E[|x_\mathbf{r}-H\mathbf{w}^H\mathbf{y}_\mathbf{r}|^2]
    \label{eqn:wiener_postfilter}\\
    &= \frac{\sigma^2_x}{\sigma^2_x + \mathbf{w}^H\Gamma_\mathbf{r}\mathbf{w}}.
\end{align}
Assuming white Gaussian noise ($\Gamma_\mathbf{r} = \sigma^2_n \mathbf{I}$, and $x \sim \mathcal{N}(0,\sigma^2_x)$ the Wiener beamformer is equivalent to Wiener post-filtering for DAS, given by:
\begin{equation}
    \hat{w}_\text{wiener} = H_\text{wiener} \mathbf{w}_\text{DAS} = \frac{\sigma^2_x}{C\sigma^2_x+\sigma^2_n}\mathbf{1}.
    \label{eqn:wiener_das}
\end{equation}

\subsubsection*{Coherence Factor weighing}
\label{sec:coherence}

The Coherence Factor (CF) \citep{Raoul1994} aims to quantify the coherence of the back-scattered echoes in order to improve image quality through scaling with a so-called coherence factor, defined as
\begin{equation}
    \text{CF} = \frac{|\mathbf{1}^H\mathbf{y}_\mathbf{r}|^2}{C\mathbf{y}^H_\mathbf{r}\mathbf{y}_\mathbf{r}},
    \label{eqn:CF}
\end{equation}
where $C$ denotes the number of channels. Effectively, this operates as a post-filter, after beamforming, based on the ratio of coherent and incoherent energy across the array. As such, it can suppress focusing errors that may occur due to speed-of-sound inhomogeneity, given by

\begin{equation}
    \hat{x}_\text{CF} = \text{CF} \cdot \hat{x}_\text{DAS} = \frac{\text{CF}}{C} \mathbf{1}^H \mathbf{y}_\mathbf{r} = \frac{\sigma^2_x}{C\sigma^2_x+C\sigma^2_n}\mathbf{1}.
    \label{eqn:CF_solution}
\end{equation}

The CF has been reported to significantly improve contrast, especially in regions affected by phase distortions. However it also suffers from reduced brightness, and speckle degradation. An explanation for this can be found when comparing \eqref{eqn:CF_solution} with the Wiener post-filter for DAS in \eqref{eqn:wiener_das}. We can see that CF weighing is in fact a Wiener post-filter where the noise is scaled by a factor $C$, leading to a stronger suppression of interference, but consequently also reducing brightness. Several derivations of the CF have been proposed to overcome some of these limitations, or to further improve image quality, such as the Generalized CF \citep{Li2003}, and Phase Coherence factor \citep{Camacho2009}.

\subsubsection*{Iterative MAP beamforming:}

\citet{Chernyakova2019}, propose an iterative maximum a posteriori (iMAP) estimator, which provides a statistical interpretation to post-filtering. The iMAP estimator works under the assumption of knowledge on the received signal model, and treats signal of interest and interference as uncorrelated Gaussian random variables with variance $\sigma^2_x$.
Given the likelihood model in \eqref{eqn:das_likelihood}, and $x \sim \mathcal{N}(0,\sigma^2_x)$, the MAP estimator of x is given by

\begin{equation}
    \hat{x}_{\text{MAP}} = 
    \frac{\sigma^2_x}{\sigma^2_n+C\sigma^2_x} \mathbf{1}^H\mathbf{y}_\mathbf{r} = 
    \frac{C\sigma^2_x}{\sigma^2_n+C\sigma^2_x}\hat{x}_\text{DAS}.
    \label{eqn:MAP_beamforming}
\end{equation}
However, the parameters $\sigma^2_x$ and $\sigma^2_n$ are unknown in practice. Instead, these can be estimated from the data at hand, leading to an iterative solution

First an initial estimate of the signal and noise variances is calculated through
\begin{equation}
    \{\hat{\sigma}^2_x, \hat{\sigma}^2_n\}_{(t)} = \left\{\hat{x}^2_{(t)}, \frac{1}{C}||\mathbf{y}_\mathbf{r}-\mathbf{1}\hat{x}_{(t)}||^2\right\},
    \label{eqn:imap_rule1}
\end{equation}
and initializing with the DAS estimate $\hat{x}_{(0)} = \frac{1}{C}\mathbf{1}^\textrm{H} \mathbf{y}$.
Following \eqref{eqn:MAP} and \eqref{eqn:imap_rule1}, a MAP estimate of the beamformed signal is given by
\begin{equation}
    \mathbf{\hat{x}}_{\text{iMAP},(t+1)} = \frac{\hat{\sigma}^2_{x,(t)}}{C\hat{\sigma}^2_{x,(t)} + \hat{\sigma}^2_{n,(t)}} \mathbf{1}^\textrm{H}\mathbf{y}_\mathbf{r}.
    \label{eqn:imap_rule2}
\end{equation}
where $t$ is an index denoting the number of iterations. Equations~\eqref{eqn:imap_rule1} and \eqref{eqn:imap_rule2} are iterated until a stopping criterion is met. The authors suggest that 2 iterations yield a sufficient noise reduction (80-100dB) for conventional ultrasound images.
If we compare \eqref{eqn:wiener_das} with \eqref{eqn:imap_rule2}, we can see that for the estimates given by \eqref{eqn:imap_rule1}, Wiener beamforming coincides with a single iteration of the iMAP algorithm.

\subsubsection*{ADMIRE}
\label{sec:ADMIRE}

Byram \textit{et al.} propose Aperture Domain Model Image Reconstruction (ADMIRE)  \citep{Byram2015, Byram2017_admire}, which explicitly models the underlying acoustic `sources' in the scan region. Let $\mathbf{A}$ be a model matrix with predictors representing both scattering from cluttered regions and scattering from the direction of interest, and $\bm{\beta}$ a vector of model parameters. We can then write the received signal at each time instance as $\mathbf{y} = \mathbf{A}\bm{\beta}$, where $\mathbf{y}$ is a single frequency bin of the received signal from a given short-time Fourier transform (STFT) window. Exact details on $\mathbf{y}$, $\bm{\beta}$ and $\mathbf{A}$ are given in \citet{Byram2015}. Performing MAP estimation for $\bm{\beta}$ under a (white) Gaussian measurement model and a mixed Gaussian/Laplacian prior yields
\begin{equation}
    \hat{\bm{\beta}} = \argmin_{\bm{\beta}} (||\mathbf{y}-\mathbf{A}\bm{\beta}||_2^2 + \lambda(\alpha ||\bm{\beta}||_1 + (1-\alpha)||\mathbf{y}||^2_2/2)),
    \label{eqn:ADMIRE}
\end{equation}
where $\alpha$ and $\lambda$ are regularization parameters. This particular form of regularization is also called elastic-net regularization. 
ADMIRE shows significant reduction in clutter due to multi-path scattering, and reverberation, resulting in a 10-20dB improvement in CNR.

\subsection*{Sparse coding}
\label{sec:sparsecoding_bf}
Chernyakova \textit{et al.} propose to formulate the beamforming process as a line-by-line recovery of back-scatter intensities from (potentially under-sampled) Fourier coefficients \citep{Chernyakova2014}. Denoting the axial fast-time intensities by $\mathbf{x}\in R^{N}$, and the noisy measured DFT coefficients of a scan line by $\tilde{\mathbf{y}}\in R^{M}$, with $M\leq N$, we can formulate the following linear measurement model:
\begin{equation}
    \tilde{\mathbf{y}} = \mathbf{H}\mathbf{F}_u\mathbf{x} + \mathbf{n} = \mathbf{A}\mathbf{x} + \mathbf{n},
\end{equation}
where $\mathbf{F}_u$ is an $M\times N$ (partial) DFT matrix, $\mathbf{H}$ a diagonal $M\times M$ matrix representing the DFT coefficients of the transmit pulse, and $\mathbf{n}$ is a white Gaussian noise vector. Recovery of $\mathbf{x}$ under a Laplace prior distribution (i.e. assuming it is sparse) can again be posed as a MAP estimation problem:
\begin{align}
    \hat{\mathbf{x}} =& \argmax_\mathbf{x} p(\tilde{\mathbf{y}}|\mathbf{x})p(\mathbf{x}) \nonumber \\
     =& \argmax_\mathbf{x} ||
    \tilde{\mathbf{y}} - \mathbf{A}\mathbf{x} ||_2^2 + \lambda||\mathbf{x}||_1,
    \label{eq:FD_coding}
\end{align}
where $\lambda$ is a regularization parameter. Problem \eqref{eq:FD_coding} can be solved using the Iterative Shrinkage and Thresholding Algorithm (ISTA), a proximal gradient method:
\begin{equation}
    \begin{aligned}
        \mathbf{\hat{x}}^{k+1} &= \mathcal{\tau}_{\lambda}(\mathbf{x}^{k}- \mu \mathbf{A}^{H}(\mathbf{A}\mathbf{x}^{k} - \tilde{\mathbf{y}})),
    \end{aligned}
    \label{eq:ISTA_beamforming}
\end{equation}
where $\mathcal{\tau}_{\lambda} = \sgn(x_{i})(|x_{i}| - \lambda)_{+} $ is the proximal operator of the $\ell_1$ norm, $\mu$ is the gradient step size, and $(\cdot)^H$ denotes the Hermitian, or conjugate transpose. It is interesting to note that the first step in the ISTA algorithm, given by $\hat{\mathbf{x}}^{1}= \mathbf{A}^H \tilde{\mathbf{y}} = \mathbf{F}^{H}_u \mathbf{H}^{H}\tilde{\mathbf{y}}$, thus maps $\tilde{\mathbf{y}}$ back to the axial/fast-time domain through the zero-filled inverse DFT. 

\subsection*{Wavefield inversion}
\label{sec:wavefield_inversion}

The previously described beamforming methods all build upon measurement models that treat pixels or scan lines (or, for ADMIRE: short-time windows) independently. As a result, complex interaction of contributions and interference from the full lateral field of view are not explicitly modeled, and often approximated through some noise model. To that end, several works explore reconstruction methods which model the joint across the full field of view, and its intricate behavior, at the cost of a higher computational footprint. 

Such methods typically rely on some form of ``wavefield inversion'', i.e. inverting the physical wave propagation model. One option is to pose beamforming as a MAP optimization problem through a likelihood model that relates the per-pixel back-scatter intensities to the channel signals \citep{szasz2016elastic, szasz2016beamforming, ozkan2017inverse}, and some prior/regularization term on the statistics of spatial distributions of back-scatter intensities in anatomical images. Based on the time-delays given by \eqref{eqn:tof_delays} (and the Green's function of the wave equation), one can again formulate our typical linear forward model:
\begin{equation}
    \mathbf{y} = \mathbf{A}\mathbf{x}+\mathbf{n}
    \label{eqn:inverse_beamforming}
\end{equation}
where $\mathbf{x} \in \mathbb{R}^{r_x r_z}$ is a vector of beamformed data, $\mathbf{n} \in \mathbb{R}^{C N_t}$ an additive white Gaussian noise vector, and $\mathbf{y} \in \mathbb{R}^{C N_t}$ the received channel data.
The space-time mapping is encoded in the sparse matrix $\mathbf{A} \in \mathbb{R}$. 

Solving this system of equations relies heavily on priors to yield a unique and anatomically-feasible solution, and yields the following MAP optimization problem:
\begin{equation}
    \hat{\mathbf{x}} = \argmin_\mathbf{x}\norm{\mathbf{y}-\mathbf{A}\mathbf{x}}_{2}^{2} - \log p_{\theta}(\mathbf{x}),
    \label{eq:formal_problem}
\end{equation}
where $\log p_{\theta}(\mathbf{x})$ acts as a regularizer, with parameters $\theta$ (e.g. an $l_{1}$ norm to promote a sparse solution \citep{combettes2005signal}).
\citet{ozkan2017inverse} investigate several intuition- and physics-based regularizers, and their effect on the beamformed image. The results show benefits for contrast and resolution for all proposed regularization methods, however each yielding different visual characteristics. 
This shows that choosing correct regularization terms and parameters that yield a robust beamformer can be challenging.

\subsection*{Post processing}
\label{sec:post_processing}

After mapping the channel data to the image domain via beamforming, ultrasound systems apply several post processing steps. Classically, this includes further image processing to boost B-mode image quality (e.g. contrast, resolution, de-speckling), but also spatio-temporal processing to suppress tissue clutter and to estimate motion (e.g. blood flow). Beyond this, we see increasing attention for post-processing methods dedicated to advanced applications such as super-resolution ultrasound localization microscopy (ULM). We will now go over some of the model-based methods for post processing, covering B-mode image quality improvement, tissue clutter filtering, and ULM. 

\subsubsection*{B-mode image quality improvement}
\label{sec:model_based_IQboost}

Throughout the years, many B-mode image-quality boost algorithms have been proposed with aims that can be broadly categorized into: 1) resolution enhancement, 2) contrast enhancement, and 3) speckle suppression. Although our focus lies on model-based methods (to recall: methods that are derived from models and first principles), it is worth nothing that B-mode processing often also relies on heuristics to accommodate e.g. user preferences. These include fine-tuned brightness curves (S-curve) to improve perceived contrast.  

A commonly used method to boost image quality is to coherently compound multiple transmissions with diverse transmit parameters. Often, a simple measurement model similar to that in DAS is assumed, where multiple transmissions are (after potential TOF-alignment) assumed to measure the same tissue intensity for a given pixel, but with different Gaussian noise realizations. As for the Gaussian likelihood model for DAS, this then simply yields averaging of the individual measurements (e.g different plane wave angles, or frequencies). More advanced model-based compounding methods use MV-weighting of the transmits, thus assuming a likelihood model where multiple measurements have correlated noise:
\begin{align}
    \hat{x}_\mathbf{r} &= \argmax_{x_\mathbf{r}} \log p(\mathbf{y}_\mathbf{r}| x_\mathbf{r},\mathbf{\Gamma}_\mathbf{r}) \\
    &= \argmin_{x_\mathbf{r}} (\mathbf{y}_\mathbf{r}-\mathbf{1}x_\mathbf{r})^H\mathbf{\Gamma}_\mathbf{r}^{-1}(\mathbf{y}_\mathbf{r}-\mathbf{1}x_\mathbf{r}). \label{eqn:MV_compounding}
\end{align}
Note that here, unlike in MV beamforming, $\mathbf{y}_\mathbf{r}$ is a vector containing the beamformed pixel intensities from multiple transmits/measurements (after TOF alignment), $\hat{x}_\mathbf{r}$ is the compounded pixel, and $\mathbf{\Gamma}_\mathbf{r}$ is the auto-correlation matrix across the series of transmits to be estimated.
Compounding can boost resolution, contrast, and suppress speckle. 

After compounding, additional processing is performed to further boost image quality. For denoising/speckle-suppression, many advanced methods have been developed, that all aim to reduce inter-tissue variations while preserving sharpness of edges. This includes directional median filtering \citep{czerwinski1995ultrasound}, adaptive kernel sizes \citep{nugroho2019artifact}, or more recently non-local-means (NLM) filters and their Bayesian extension, the Optimized Bayesian NLM (OBNLM) filter. These methods formulate denoising as a patch-based MAP inference problem \citep{kervrann2007bayesian}, where an unobserved image patch $\mathbf{x}$ with some unknown probability density function is estimated from its noisy observation $\mathbf{y}=f(\mathbf{x}+\mathbf{n})$, with $f(\cdot)$ denoting a function related to the image formation processes, and $\mathbf{n}$ a random noise vector with independent and identically distributed (i.i.d.) entries. The density functions are then estimated by exploiting redundancy across patches in an image, drawing samples from a local neighborhood. This probabilistic Bayesian interpretation to NLM has enabled ultrasound-specific implementations with more realistic (multiplicative) noise models \citep{coupe2008bayesian}. Other MAP approaches pose denoising as a dictionary matching problem \citep{jabarulla2018speckle}. These methods do not explicitly estimate patch density functions from the image, but instead learn a dictionary of patches. Other approaches combine spatial filters with priors (e.g. sparsity) in a transformed domain, i.e., the wavelet domain \citep{garg2019despeckling}, or combine PCA and wavelet transforms \citep{jagadesh2016novel}.

To achieve a boost in image resolution, the problem can be recast as MAP estimation under a likelihood model that includes a deterministic blurring/point-spread-function matrix $\mathbf{A}_{\text{blur}}$:
\begin{equation}
    \hat{\mathbf{x}} = \argmin_\mathbf{x}\norm{\mathbf{y}-\mathbf{A}_{\text{blur}}\mathbf{x}}_{2}^{2} - \log p_{\theta}(\mathbf{x}),
    \label{eq:deblurring}
\end{equation}
where $\mathbf{x}$ is the (vectorized) high-resolution image to be recovered, and $\mathbf{y}$ a (vectorized) blurred and noisy (Gaussian white) observation $\mathbf{y}$. This deconvolution problem is ill posed and requires adequate regularization via priors. As we noted before, the log-prior term can take many forms, including $\ell_1$ or total variation based regularizers.

\subsubsection*{Clutter filtering for flow}

For many applications (including flow mapping and localization microscopy), suppression of reflections from tissue is of interest \cite{Wildeboer2020blind}. Slow moving tissue introduces a clutter signal that introduces artefacts and obscures the feature of interest being imaged (be it blood velocity or e.g. contrast agents), and considerable effort has gone into suppressing this tissue clutter signal. 
Although Infinite Impulse Response (IIR) and Finite Impulse Response (FIR) filters have been the most commonly used filters for tasks such as this, it is still very difficult to separate the signals originating from slow moving blood or fast moving tissue. Therefore, spatio-temporal clutter filtering is receiving increasing attention. We will here go over some of these more advanced methods (including singular value thresholding and robust principle component analysis), again taking a probabilistic MAP perspective. 

We define the spatio-temporal measured signal as a Casorati matrix, $\mathbf{Y} \in \mathbb{R}^{NM \times T}$, where $N$ and $M$ are spatial dimensions, and $T$ is the time dimension, which we model as $\mathbf{Y} = \mathbf{X}_{\text{tissue}} + \mathbf{X}_{\text{blood}}$, where $\mathbf{X}_{\text{tissue}} \in \mathbb{R}^{NM \times T}$ is the tissue component, and $\mathbf{X}_{\text{blood}} \in \mathbb{R}^{NM \times T}$ is the blood/flow component. 
We then impose a prior on $\mathbf{X}_{\text{tissue}}$, and assume it to be \textit{low rank}. If we additionally assume $\mathbf{X}_{\text{blood}}$ to have i.i.d. Gaussian entries, the MAP estimation problem for the tissue clutter signal becomes:
\begin{equation}
    \begin{aligned}
        \hat{\mathbf{X}}_\text{\text{tissue}} &= \argmax_{\mathbf{X}_\text{\text{tissue}} } p(\mathbf{Y}|\mathbf{X}_{\text{tissue}}) p(\mathbf{X}_{\text{tissue}}) \\
        &= \argmax_{\mathbf{X}_{\text{tissue}}}\log p(\mathbf{Y}|\mathbf{X}_{\text{tissue}}) +\log(p(\mathbf{X}_{\text{tissue}})) \\
        &= \argmin_{\mathbf{X}_{\text{tissue}}} \norm{\mathbf{Y}-\mathbf{X}_{\text{tissue}}}_F + \lambda \norm{\mathbf{X}_{\text{tissue}}}_{*},
    \end{aligned}
    \label{eqn:SVD}
\end{equation}
where $\norm{\cdot}_{F}$ and $\norm{\cdot}_{*}$ denote the Frobenius norm and the nuclear norm, respectively. The solution to \eqref{eqn:SVD} is:
\begin{equation}
    \hat{\mathbf{X}}_{\text{tissue}} = \mathcal{T}_{\text{SVT},\lambda}(\mathbf{Y}),
\end{equation}
where $\mathcal{T}_{\text{SVT},\lambda}$ is the singular value thresholding function, which is the proximal operator of the nuclear norm \citep{cai2010singular}.

To improve upon the model in \eqref{eqn:SVD}, one can include a more specific prior on the flow components, and separate them from the noise:
\begin{equation}
    \mathbf{Y} = \mathbf{X}_{\text{tissue}} + \mathbf{X}_{\text{blood}} + \mathbf{N},
\end{equation}
where we place a mixed $\ell_1/\ell_2$ prior on the blood flow component $\mathbf{X}_{\text{blood}}$, and assume i.i.d. Gaussian entries in the noise matrix $\mathbf{N}$, such that: 
\begin{equation*}
        \begin{aligned}
        \hat{\mathbf{X}}_{\text{tissue}}, \hat{\mathbf{X}}_{\text{blood}} 
        &= \argmax_{\mathbf{X}_{\text{tissue}}, \mathbf{X}_{\text{blood}}}p(\mathbf{Y}|\mathbf{X}_{\text{tissue}}, \mathbf{X}_{\text{blood}})p(\mathbf{X}_{\text{tissue}}) \\
        & \qquad p(\mathbf{X}_{\text{blood}}) \\
        &= \argmax_{\mathbf{X}_{\text{tissue}}, \mathbf{X}_{\text{blood}}}\log p(\mathbf{Y}|\mathbf{X}_{\text{tissue}}, \mathbf{X}_{\text{blood}}) + \\
        & \qquad \log(p(\mathbf{X}_{\text{tissue}})) +\log(p(\mathbf{X}_{\text{blood}})) \\ &= \argmin_{\mathbf{X}_{\text{tissue}}, \mathbf{X}_{\text{blood}}} \norm{\mathbf{Y}-\mathbf{X}_{\text{tissue}}-\mathbf{X}_{\text{blood}}}_{F} + \\
        & \qquad \qquad \lambda_{1} \norm{\mathbf{X}_{\text{tissue}}}_{*} + \lambda_{2} \norm{\mathbf{X}_{\text{blood}}}_{1,2}
    \end{aligned}
\end{equation*}
where $\norm{\cdot}_{1,2}$ indicates the $\ell_{1}$ and $\ell_{2}$ norm. This \textit{low-rank plus sparse} optimization problem is also termed Robust Principle Component Analysis (RPCA), and can be solved through an iterative proximal gradient method:
\begin{equation}
    \begin{aligned}
        \mathbf{X}_{\text{tissue}}^{k+1} =& \mathcal{T}_{\text{SVT},\lambda_1}(\mathbf{X}_{\text{tissue}}^{k} - \mu_1 (\mathbf{Y} - \mathbf{X}_{\text{tissue}}^{k} - \mathbf{X}_{\text{blood}}^{k}))
    \label{eqn:rpca_tis}
    \end{aligned}
\end{equation}

\begin{equation}
    \begin{aligned}
        \mathbf{X}_{\text{blood}}^{k+1} =& \mathcal{T}_{1,2,\lambda_2}(\mathbf{X}_{\text{blood}}^{k} - \mu_2 (\mathbf{Y} - \mathbf{X}_{\text{tissue}}^{k} - \mathbf{X}_{\text{blood}}^{k})),
    \label{eqn:rpca_bld}
    \end{aligned}
\end{equation}
where $\mathcal{T}_{\text{SVT},\lambda_1}$ is the solution of \eqref{eqn:SVD} (i.e. the proximal operator of the nuclear norm), $\mathcal{T}_{1,2,\lambda_1}$ is the mixed $\ell_1$-$\ell_2$ thresholding operation, and $\mu_1$ and $\mu_2$ are the gradient steps for the two terms.

\citet{shen2019high} further augment the RPCA formulation to boost resolution for the blood flow estimates. To that end they add a PSF-based convolution kernel to the blood component $\mathbf{A}_{r} \circledast \mathbf{X}_{blood}$, casting it as a joint deblurring and signal separation problem. 

\subsubsection*{Ultrasound Localization Microscopy}
\label{sec:model_based_ulm}
We will now turn to an advanced and increasingly popular ultrasound signal processing application: ULM. Conventional ultrasound resolution is fundamentally limited by wave physics, to half the wavelength of the transmitted wave, i.e., the diffraction limit. This limit is in the range of millimeters for most ultrasound probes, and is inversely proportional to the transmission frequency. However, high transmit frequencies come at the cost of lower penetration depth. 

To overcome this diffraction limit, ULM adapts concepts from Nobel-prize winning super-resolution fluorescence microscopy to ultrasound. Instead of localizing fluorescent blinking molecules, ULM detects and localizes ultrasound contrast agents, microbubbles, flowing through the vascular bed. These microbubbles have a size similar to red blood cells, and act as point scatterers. By accumulating precisely localized microbubbles across many frames, a super-resolution image of the vascular bed can be obtained. In typical implementations, the localization of the MB's is performed by centroid detection \citep{Siepmann2011, Couture2011, christensen2020super}.

Not surprisingly, we can also pose microbubble localization as a MAP estimation problem \citep{van2017sparsity}. We define a sparse high-resolution image that is vectorized into $\mathbf{x}$, in which only few pixels have non-zero entries: those pixels that contain a microbubble. Our vectorized measurements can then be modeled as: $\mathbf{y}=\mathbf{A}\mathbf{x}+\mathbf{n}$, where $\mathbf{A}$ is a PSF matrix and $\mathbf{n}$ is a white Gaussian noise vector. This yields the following MAP problem:
\begin{equation}
    \begin{aligned}
        \hat{\mathbf{x}} &= \argmax_{\mathbf{y}} p(\mathbf{y}|\mathbf{x})p(\mathbf{x}) \\
        &= \argmax_{\mathbf{x}}\log p(\mathbf{y}|\mathbf{x}) +\log p(\mathbf{x}) \\
        &= \argmin_{\mathbf{x}} \norm{\mathbf{y}-\mathbf{A}\mathbf{x}}_{2}^{2} + \lambda \norm{\mathbf{x}}_{1}.
    \end{aligned}
\end{equation}
\citet{van2017sparsity} propose to solve this sparse coding problem using ISTA, similar to the formulation in \eqref{eq:ISTA_beamforming}.

Instead of processing each image frame independently, Bar-Zion \textit{et al.}  exploit sparse structure in the temporal correlation domain, i.e. $\mathbf{y}$ is a correlation image \citep{bar2018sushi}, leading to the SUSHI method. Later, Solomon \textit{et al}. combine MAP estimation across the spatial dimensions with MAP estimation in time, by complementing the spatial sparse coding problem with a Kalman filter that places a prior on future microbubble locations according to a motion model \citep{solomon2019exploiting}. 

\section*{Deep Learning for US Signal Processing}\label{sec:data_driven}

\noindent Deep learning based ultrasound signal processing offers a highly flexible framework for learning a desired input-output mapping $\hat{X} = f_\theta(Y)$ from training data, overcoming the need for explicit modeling and derivation of solutions. This can especially be advantageous for complex problems in which models fall short (e.g. incomplete, with naive assumptions) or their solutions are demanding or even intractable. We will now go over some emerging applications of deep learning in the ultrasound signal processing pipeline. As in the previous section, we will first cover advanced methods for beamforming and then turn to downstream post-processing such as B-mode image quality improvement, clutter suppression and ULM.

\subsection*{Beamforming}
\noindent We discern two categories of approaches: Neural networks that replace the entire mapping from channel data to images, and those that only replace the beamsumming operation, i.e. after TOF correction. \citet{hyun2021deep} and \citet{bell2020challenge, bell2019challenge19} recently organized the Challenge on Ultrasound Beamforming with Deep Learning (CUBDL) to incentivise new research in this area. For a more in-depth survey on deep learning for ultrasound beamforming, including common training strategies and loss functions, we refer the reader to \citet{van2021deep}.

\subsubsection*{Direct channel to image transformation}
The authors in \citet{nair2018deep, Nair2020} introduce a method that learns a direct convolutional neural network based transformation between the channel signals and a target B-mode image. The proposed U-net architecture thus has to learn both the (geometry-based) time-of-flight correction, as well as the subsequent beamsumming. The inputs and outputs of the network comprise IQ-demodulated channel data (separate I and Q inputs)  $\mathbf{Y} \in \mathbb{R}^{C \times N_t \times 2}$, and a beamformed image $\mathbf{X} \in \mathbb{R}^{R_x \times R_y}$, respectively. The network additionally outputs direct segmentations of anechoic regions in the image. Training is done using ultrasound simulations of a variety of anechoic cysts. 

Replacing the entire beamforming pipeline is an unconventional application of U-net-style architectures, which were originally designed for image-to-image translation (segmentation) problems. Instead, the U-net here also performs time-to-space migration. It is worth noting that much of the efficiency of convolutional networks comes from their ability to exploit spatial symmetries (i.e. translation/shift equivariance): their operations do not depend on the position in the data, i.e. they are spatially invariant. In contrast, TOF correction is based on the geometry of the array and the scan region, and its operation varies depending on the focus point. As such, learning time-space migration through standard convolutional architectures is not trivial. Most beamforming approaches thus benefit from traditional alignment of the channel data before processing. 

\subsubsection*{Beam-Summing after TOF correction}

While the solution proposed by Nair \textit{et al.} replaces the entire beamformer with a convolutional neural network (including the time-space migration), most works confine the problem to only the beamsumming and compounding steps, leaving the TOF-correction based on geometry. As such it replaces the model-based probabilistic beamforming/summing methods, mapping TOF-corrected channel data $\mathbf{Y} \in \mathbb{R}^{C \times R_x \times R_y}$, to an image $\mathbf{X} \in \mathbb{R}^{R_x \times R_y}$. In advanced adaptive beamforming with intricate models this step is often the most time-consuming and complex, in some cases preventing real-time implementation. Often, an important goal of DL-based beamsumming is therefore to accelerate these advanced beamsumming methods by learning fast neural network approximations. Many of the methods we will list here make use of deep convolutional architectures, which have a significant spatial receptive field: the summed output at a given pixel is based on a large spatial input support. This spatial `awareness' is in contrast with most of the model-based beamforming methods, that operate on a per-pixel or sometimes per-line basis. 

In the work by \citet{khan2019universal,khan2020adaptive}, a deep learning based beamforming network is proposed which replaces the conventional summation of channel signals and envelope detection by a deep CNN. Additionally the authors show that their method enables image reconstruction from sub-sampled RF data, in which a set of receive elements is inactive, hereby reducing the required bandwidth.
In an extension of this work \citet{khan2020switch} show that such a beamformer can be optimized for different imaging settings, and controlled through the introduction of a style code vector in the latent space and training on a corresponding image target with a given style/setting. Such an approach avoids the need to store separate models for each setting.

\citet{Vignon2020} propose a similar solution as Khan \textit{et al.}, in which line-wise channel signals are beamsummed by a CNN. It is worth noting that in this work training data is generated using purely simulations, in which targets correspond to DAS beamformed images obtained with larger simulated array aperture to yield better imaging resolution.

\citet{mamistvalov2021deeplearning} propose a U-net architecture for beamsumming after Fourier-domain TOF correction.
This enables sub-Nyquist acquisition by Xampling \citep{Chernyakova2014}, reducing data rates. The authors show that their convolutional U-net architecture enables reconstruction of high-quality images from fast-time sub-Nyquist acquisitions acquired with a sparse array (channel subsampling) \citep{Cohen2018}, suppressing the aliasing artifacts. 

Similarly, \citet{huijben2020learning} perform image reconstruction from sparse arrays of undersampled channels using a convolutional network architecture. In addition, the authors provide a mechanism for jointly learning optimal channel selection/sparse array design via a technique dubbed deep probabilistic subsampling.

While most beamforming methods aim at boosting resolution and contrast, \citet{hyun2019beamforming} argue that beamformers should accurately estimate the true tissue backscatter map, and thus also target speckle reduction. The authors train their beamformer on ultrasound simulations of a large variety of artificial tissue backscatter maps derived from natural images.

\subsection*{Post processing}
Application of deep learning methods to general image processing/restoration problems has seen a surge of interest in recent years, showing remarkable performance across a range of applications. Naturally, these pure image processing methods are being explored for ultrasound post processing as well. In this section we will treat the same topics as in the previous chapter, but focus on recent deep learning methods. 

\subsubsection*{B-mode image quality improvement}
A common means to boost ultrasound image quality is compounding data from different transmits. While model-based methods offer simple pixel-based compounding strategies (either by simply summing or via MV processing), several groups have investigated the use of deep CNNs for improved compounding \citep{khan2019universal, lu2019fast, guo2020high}. \citet{jansen2021enhanced} propose to perform this compounding step in the Radon domain.

In addition to compounding multiple transmits, many deep learning methods aim at single-image enhancement, including resolution/contrast boost, but also speckle suppression. \citet{gasse2017high} explore mapping a single PW image to an image that was compounded using multiple PWs. Zhou \textit{et al}. pursue a similar goal, but introduce a multi-branch CNN with an additional wavelet-based post-processing step \citep{zhou2018high}. \citet{qi2020image} perform processing in the Fourier domain, using pixel-wise fully connected neural networks. Rather than compounded PWs, the authors use a focused line-scan image as a target. 

The work by \citet{chang2019two} poses denoising as a signal decomposition problem. To that end they propose a two-stage CNN that simultaneously models the image and noise, where the noise estimates in turn inform the image estimates to cope with various noise distributions.

\citet{temiz2020super} aim at single-image super-resolution, i.e. to achieve a B-mode image with a higher pixel resolution. The authors achieve this by training a deep CNN with a dataset containing B-mode US images across a range of scale/zoom factors. A similar approach was taken by \citet{choi2018deep}, where they propose a deep CNN called SRGAN with the aim to map low resolution images to a high resolution domain.

Both \citet{vedula2017towards} and \citet{ando2020speckle} approach the issue of speckle reduction is similar ways, by using a CNN. However, they use different forms of input and target data, while \citet{vedula2017towards} use IQ data, \citet{ando2020speckle} uses B-mode images. Similarly, \citet{dietrichson2018ultrasound} perform this task by using a CNN, albeit with a more complex training strategy, by employing a GAN based structure. \citet{karaouglu2021removal} compare many approaches to this problem from the perspective of neural networks, and detail the effectiveness of the many architectures. They find that the GAN, and the U-Net like algorithms in their study performed the best. While all the works cited on this topic so far deal with 2D US scans, \citet{li20203d} attempt to extend it to 3D imaging using a 3D UNet model. It is an interesting point to note that \citet{vedula2017towards} and \citet{ando2020speckle} use simulated data, while the other works on speckle reduction use \textit{in-vivo} data gathered from volunteers. However, there is uniformity in how these works create their target images; through model based speckle reduction algorithms.

Most deep learning methods for image quality improvement rely on supervised learning, requiring ground truth targets which are often difficult to obtain. As an alternative, \citet{huh2021tunable} present a self-supervised method based on the cycle-GAN architecture, originally developed for unpaired (cycle-consistent) style transfer \citep{huh2021tunable}. This approach aims at transferring the features of a high-quality target distribution of images to a given low-quality image, which the authors leverage to improve elevational image quality in 3D ultrasound.

\subsubsection*{Clutter filtering for flow}
\citet{brown2020deep} describe a 3D (2D + time) CNN-based spatio-temporal filtering scheme for tissue clutter suppression, mostly aimed at accelerating the SVD algorithm for real-time use. To that end, they use SVD processed \textit{in-vivo} and \textit{in-vitro} images as targets. Similarly, Wang \textit{et al.} aim at replacing SVD thresholding \citep{wang2021preliminary}, but rather than a 3D CNN, the authors adopt a 2D (spatial) CNN, aggregating temporal information in the feature space through a recurrent neural network.

\citet{tabassian2019clutter} use a deep 3D CNN (2D + time) to suppress clutter and reverberation artifacts that plague echocardiographic imaging. Their deep network was trained on realistic simulations of echocardiographic exams, with simulated superimposed artifacts.

\subsubsection*{Ultrasound Localization Microscopy}
\citet{van2019sr,van2020super} propose a deep learning method based on a encoder-decoder architecture that aims to replace costly iterative techniques for sparse coding to obtain super-resolution vascular images from high-density contrast-enhanced ultrasound data. Later, \citet{liu2020deep} propose a similar approach, but use a sub-pixel CNN. Brown \textit{et al.} propose to jointly perform tissue clutter filtering and localization by a 3D CNN to further boost processing rates \citep{brown2021faster}. Notably, \citet{youn2020detection} perform localization directly from channel data. 

\section*{Model-Based Deep Learning for US Signal Processing}\label{sec:model_based_dl}

We now highlight several works that incorporate signal processing knowledge in their deep learning approaches to improve performance, reduce network complexity, and to provide reliable inference models. Generally, these models retain a large part of the conventional signal processing pipeline intact, and replace critical points in the processing with neural networks, so as to provide robust inference as a result. We will discuss methods ranging from iterative solvers, to unfolded fixed complexity solutions. 

\subsection*{Beamforming}

\subsubsection*{Model-based pre-focussing using DL}

Pre-focussing (or TOF correction) is conventionally done deterministically, based on the array geometry and assuming a constant speed-of-sound.
Instead, data-adaptive focusing, by calculating delays based on the recorded data, facilitates correction for speed-of-sound mismatches.
The work by \citet{nair2018deep,Nair2020} does this implicitly, by finding a direct mapping from the time-domain to an output image, using DL. However, this yields a black-box solution, which can be difficult to interpret. 

The authors of \citet{YoungMin2021} adhere more strictly to a conventional beamforming structure, and tackle this problem in two steps: first the estimation of a local speed-of-sound map, and secondly the calculation of the corresponding beamforming delays. The speed-of-sound image is predicted from multi-angled plane wave transmissions using SQI-net \citep{SeokHwan2021}, a type of U-net. One then needs to find the propagation path and travel time of the transmitted pulse, i.e. the delay-matrix, between each imaging point and transducer element.
For a uniform speed-of-sound this is trivial, since the shortest distance between a point and element corresponds to the fastest path. For a non-uniform speed-of-sound, this is more challenging, and requires a path finding algorithms that adds to the computational complexity. The Dijkstra algorithm \citep{Dijkstra1959note} for instance, which is commonly used to find the fastest path, has a complexity of $\mathcal{O}(n^2 \log n)$, where $n$ is the number of nodes in the graph, or equivalently, the density of the local speed-of-sound grid. 

As such, the authors propose a second U-net style neural network, referred to as DelayNet, for estimating these delay times. The network comprises $3\times3$ locally masked convolutions, such that no filter weights are assigned in the direction opposite from direction of wave propagation. Intuitively, this can be understood as enforcing an increasing delay-time the further we get from the transducer, i.e. the wave does not move in reverse direction. Furthermore, the reduced filter count improves computational efficiency by $\sim$33\%.

Finally, the predicted delay matrix is used to focus the RF data using the corrected delays, after which it is beamsummed to yield a beamformed output signal. As such, DelayNet does not to be trained directly on a target delay-matrix, but instead can be trained end-to-end on the desired beamformed targets. Note that in this method, the estimation of the speed-of-sound is done in a purely data-driven fashion. However, the pre-focussing itself inherits a model-based structure, by constraining the problem to learning time-shifts from the aforementioned speed-of-sound map.

\subsubsection*{Model-based beamsumming using DL}
\label{sec:model-basedDL_BF}
\citet{luijten2019deep, luijten2020adaptive} propose adaptive beamforming by deep learning (ABLE), a deep learning based beamsumming approach that inherits its structure from adaptive beamforming algorithms, specifically minimum variance (MV) beamforming. 
ABLE specifically aims to overcome the most computationally complex part of the beamforming, the calculation of the adaptive apodization weights, replacing this with a neural network $f_\theta$. The step from the model-based MAP estimator to ABLE is then given by

\begin{align}
    \hat{x}_\mathbf{r} &= \argmax_{x_\mathbf{r}} p(\mathbf{y}_\mathbf{r}| x_\mathbf{r}) = ( \mathbf{1}^H\mathbf{\Gamma}_\mathbf{r}^{-1}\mathbf{1})^{-1} \mathbf{1}^H\mathbf{\Gamma}_\mathbf{r}^{-1}\mathbf{y}_\mathbf{r} \label{eqn:MV_mlestimator2} \\
    &= \argmax_{x_\mathbf{r}} p(\mathbf{y}_\mathbf{r}| x_\mathbf{r}) \approx f_\theta(\mathbf{y_r})^H\mathbf{y}_r, \label{eqn:ABLE}
\end{align}
where $\theta$ comprise the neural network weights, and $\mathbf{y_r}$ the TOF corrected RF data. Multiplying the predicted weights with the TOF corrected data, and summing the result, yields a beamformed output signal. 

Note that for training, we do not need access to the apodization weights as in MV beamforming. Instead, this is done end-to-end towards a MV generated target, given by
\begin{equation}
\underset{\theta}{\arg\min} \left[ \mathcal{L} \left( 
f_\theta(\mathbf{y_r})^H
\mathbf{y_r}]) - 
\mathbf{\hat{x}}_{MV}
\right]
\right),
\label{eqn:US_problem}
\end{equation}
where $\mathbf{\hat{x}}_{\text{MV}}$ is a MV training target, and $\mathcal{L}$ a loss function. 
Since the network operates directly on RF data, which has positive and negative signal components, as well as a high dynamic range, the authors propose an Antirectifier as an activation function. The Antirectifier introduces a non-linearlity while preserving the sign information and dynamic range, unlike the rectified linear unit, or hyperbolic tangent.
Similarly, a signed-mean-squared-logarithmic-error (SMSLE) loss function is introduced, which ensures that errors in the RF domain reflect the errors in the log-compressed output image. The authors show that a relatively small network, comprising four fully connected layers, can solve this task, and is able to generalize well to different datasets. They report an increase in resolution and contrast, while reducing computational complexity by 2 to 3 orders of magnitude.

\citet{wiacek2020coherenet} similarly exploit DNNs as a function approximator in order to accelerate the calculation of the short-lag spatial coherence (SLSC). 
Specifically the authors apply their method to SLSC beamforming, which displays the spatial coherence of backscattered echoes across the transducer array. This contrasts conventional DAS beamforming in which the recorded pressures are visualized.
The authors report a 3.4 times faster computation compared to the standard CPU based approach, corresponding to a framerate of 11 frames-per-second.

\citet{Luchies2018} propose a wideband DNN for suppressing off-axis scattering, which operates in the frequency domain, similar to ADMIRE discussed earlier. After focusing an axially gated section of channel data, the RF signals undergo a discrete Fourier transform (DFT), mapping the signal into different frequency bins. The neural network operates specifically on these frequency bins, after which the data is transformed back to the time-domain using the inverse discrete Fourier transform (IDFT) and summed to yield a beamformed signal. The same fully connected network structure was used for different center frequencies, only retraining the weights.

An extension of this work is described in \citet{Khan2021performingaperture}, where the neural network itself is replaced by a model-based network architecture. The estimation of model parameters $\beta$, as formulated in \eqref{eqn:ADMIRE}, can be seen as a sparse coding problem $\mathbf{y} = \mathbf{A}\beta$ (where $\beta$ is a sparse vector) which can be solved by using an iterative algorithm such as ISTA. This yields
\begin{equation}
    \begin{aligned}
        \mathbf{\hat{\beta}}^{k+1} &= \mathcal{\tau}_{\lambda}(\mathbf{\beta}^{k}- \mu \mathbf{A}^{T}(\mathbf{A}\mathbf{\beta}^{k} - \mathbf{y})),
    \end{aligned}
    \label{eqn:ista_admire}
\end{equation}
where $\mathcal{\tau}_{\lambda}(\cdot)$ is the soft-thresholding function parameterized by $\lambda$.

To derive a model-based network architecture, \eqref{eqn:ista_admire} is unfolded as a feed-forward neural network with input $\mathbf{A}^T \mathbf{y}$ and output $\hat{\beta}$, the predicted model coefficients. For each iteration, or fold, we can then learn the weight matrices, and the soft-thresholding parameter $\lambda$ trainable. This then leads to a learned ISTA algorithm (LISTA):
\begin{equation}
    \begin{aligned}
        \mathbf{\hat{\beta}}^{k+1} &= \mathcal{\tau}_{\lambda^{k}}(\mathbf{W}^{k}\mathbf{\beta}^{k} + \mathbf{A}^T\mathbf{y}),
    \end{aligned}
    \label{eqn:unfolded_admire}
\end{equation}
where $\mathbf{W}^{k}$ represents a trainable fully connected layer and $\lambda^{k}$ is a (per-fold) trainable thresholding parameter. When contrasted with its model-based iterative counterpart ISTA, LISTA is a fixed complexity solution that tailors its processing to a given dataset using deep learning. Compared to conventional deep neural networks, LISTA has a low number of trainable parameters however.

The authors show that LISTA can be trained on model fits of ADMIRE, or even simulation data containing targets without off-axis scattering, thereby potentially outperforming the fully model-based algorithm, ADMIRE, due to its ability to learn optimal regularization parameters from data.

\citet{mamistvalov2021deep} take a similar avenue and recast their model-based solution for reconstructing images from sub-Nyquist acquisitions (across channels and time) into an unfolded LISTA architecture. While the aforementioned method based on ADMIRE learns to estimate sparse codes in the aperture/channel-domain dictionary, the method by \citet{mamistvalov2021deep} learns to sparsely encode (fast-time) RF lines. The original ISTA-based algorithm is recast as a fixed-length feed forward model in which the matrix operations are replaced with learned convolutional layers. By training the network on pairs of sub-Nyquist and full Nyquist-rate data, the authors show that their approach enables a reduction in data-rates up to $\sim 88 \%$ without significantly compromising image quality. 

\subsubsection*{Model-based wavefield inversion using DL}

Reconstruction techniques based on the inversion of a (non-)linear measurement model are often very computationally intensive, and require careful tuning of hyper-parameters to ensure robust inference. Alternatively, \citet{almansouri2018deep} propose a two-step approach which leverages a simple linear model to obtain an initial estimation, after which further refinement is done through a CNN. As such, the neural network can account for non-linear, and space-varying, artifacts in the measurement model. 

The ultrasound forward model is based on a set of differential equations, and mainly depends on three parameters: the acoustic velocity $c_0$, the density $\rho_0$, and the attenuation $\alpha_0$. Such a model could abstractly be defined as
\begin{equation}
    y = f(c_0, \rho_0, \alpha_0).
\end{equation}
However, due to the complex non-linear nature of this forward model, a simplified linear model was developed in (details are given in \citet{Almansouri2018Anisotropic}), which yields the estimator
\begin{equation}
    \hat{\mathbf{x}} = \argmin_\mathbf{x}\norm{\mathbf{y}-\mathbf{A}\mathbf{x}}_{2}^{2} - \log p_{\theta}(\mathbf{x}),
    \label{eq:formal_problem_v2}
\end{equation}
where $A$ is a matrix that accounts for time-shifting and attenuation of the transmit pulse. The adjoint operator operator of the linearized model gives an approximate estimator for $x$, given by $\tilde{\mathbf{x}} = \mathbf{A}^T\mathbf{y}$. The authors adopt a U-net architecture, to compensate for artifacts caused by non-linearities. Effectively the the network finds a mapping from a relatively simple estimate, yet based on the physical measurement model, and maps it to a desired high-quality image such that
\begin{equation}
    \hat{\mathbf{x}} \approx f_\theta(\mathbf{A}^T\mathbf{y}),
    \label{eqn:almansouri}
\end{equation}
where $f(\cdot)_\theta$ denotes the neural network, and $\hat{x}$ the high-quality estimate.

\subsection*{Post-Processing and Interpretation}

\subsubsection*{Deep Unfolding for B-mode IQ enhancement/PW compounding/Compressed Acquisition}

\noindent \citet{chennakeshava2020high, chennakeshava2021deep} propose a plane wave compounding and deconvolution method based on deep unfolding. Their architecture is based on a proximal gradient descent algorithm derived from a model-based MAP optimization problem, that is subsequently unfolded and trained to compound 3 plane wave images, gathered at low frequency, into an image gathered using 75 compounded plane wave transmissions at a higher frequency. This encourages a learned proximal operator that maps low-resolution, low-contrast input images onto a manifold of images with better spatial resolution and contrast. 

Denote $\mathbf{x} \in \mathbb{R}^{N}$ as the vectorized high-resolution beamformed RF image, and $\mathbf{y} \in \mathbb{R}^{NM}$ the vectorized measurement of low-resolution beamformed RF images from $M=3$ transmitted plane waves. The authors assume the following acquisition model:
\begin{equation}
    \mathbf{y} = \mathbf{A} \mathbf{x} + \mathbf{n},
    \label{eqn:measurement_model_compounding}
\end{equation}
where
\begin{equation}
    \mathbf{A} = \begin{pmatrix}\mathbf{A}_{1} \\
    \mathbf{A}_{2} \\
    \vdots \\ \mathbf{A}_{M}\end{pmatrix},
    \label{eq:observation_m}
\end{equation}
and
\begin{equation}
    \mathbf{y} = \begin{pmatrix}\mathbf{y}_{1} \\
    \mathbf{y}_{2} \\
    \vdots \\ \mathbf{y}_{M}\end{pmatrix},
    \label{eq:measured_m}
\end{equation}
where $\mathbf{y}_{m}$ is the vectorised, beamformed RF image belonging to the $m^{\textrm{th}}$ steered plane wave transmission, $\mathbf{n}$ $\in$ $\mathbb{R}^{NM}$ is a noise vector which is assumed to follow a Gaussian distribution with zero mean and diagonal covariance, and $\mathbf{A}$ $\in$ $\mathbb{R}^{NM \times N}$ is a block matrix, with its blocks $\mathbf{A}_{1}$, $\mathbf{A}_{2}$,..., $\mathbf{A}_{M}$ being the measurement matrices of individual PW acquisitions. The authors assume that the measurement matrices (which capture the system PSF for each PW) follow a convolutional Toeplitz structure. 

Based on this model, each fold in the unfolded proximal gradient algorithm aimed at recovering the high-resolution image $\mathbf{x}$ is written as, 
\begin{equation}
    \mathbf{\hat{x}}^{(k+1)} = \mathcal{P}_{\theta}^{(k)}(\mathbf{W}^{(k)}\mathbf{y}+\mathbf{V}^{(k)}\mathbf{\hat{x}}^{(k)})
\end{equation}
where $\mathcal{P}_{\theta}$ is a Unet-style neural network replacing the generalised proximal operator, and 
\begin{equation}
    \mathbf{W}^{(k)}\mathbf{y} := \mathbf{w}_{1}^{(k)} \circledast \mathbf{y}_{1} + \mathbf{w}_{2}^{(k)} \circledast \mathbf{y}_{2} + ... + \mathbf{w}_{M}^{(k)} \circledast \mathbf{y}_{m}.
    \label{eq:seven}
\end{equation}

\begin{equation}
    \mathbf{V}^{(k)}\mathbf{\hat{x}}^{(k)} = \mathbf{v}^{(k)} \circledast \mathbf{\hat{x}}^{(k)},
    \label{eq:eight}
\end{equation}
where $\circledast$ denotes a convolutional operation, and $\lbrace\mathbf{w}_{1}^{(k)}, ..., \mathbf{w}_{m}^{(k)}\rbrace$ and $\mathbf{v}^{(k)}$ are learned convolutional kernels. The authors show that their model-based deep learning architecture outperforms model-agnostic deep learning methods, yielding high-contrast and high-resolution outputs. 

\subsubsection*{Deep unfolding for clutter filtering}
\label{sec:deep_unfolded_rpca}

\noindent \citet{solomon2019deep} propose deep unfolded convolutional robust RPCA for ultrasound clutter suppression. The approach is derived from the RPCA algorithm, given by \eqref{eqn:rpca_tis} and \eqref{eqn:rpca_bld}, but unfolds it and learns all the parameters (gradient projection and regularization weights) from data. Each network layer in the unfolded architecture takes the following form:

\begin{equation}
    \begin{aligned}
        \hat{\mathbf{X}}_{\text{tissue}}^{(k+1)} = \mathcal{T}_{\text{SVT}}^{(k)} \left( \mathbf{W}_{1}^{(k)} \circledast \mathbf{Y} + \mathbf{W}_{3}^{(k)} \circledast \mathbf{X}_{\text{blood}}^{(k)} + \right. \\
        \left. \mathbf{W}_{5}^{(k)} \circledast \mathbf{X}_{\text{tissue}}^{(k)} \right)
    \end{aligned}
\end{equation}
and,
\begin{equation}
    \begin{aligned}
        \hat{\mathbf{X}}_{\text{blood}}^{(k+1)} = \mathcal{T}_{\lambda}^{(k)} \left( \mathbf{W}_{2}^{(k)} \circledast \mathbf{Y} + \mathbf{W}_{4}^{(k)} \circledast \mathbf{X}_{\text{blood}}^{(k)} + \right. \\
        \left. \mathbf{W}_{6}^{(k)} \circledast \mathbf{X}_{\text{tissue}}^{(k)} \right)
    \end{aligned}
\end{equation}
where $\mathbf{W}_{1}, \mathbf{W}_{2}, \mathbf{W}_{3}, \mathbf{W}_{4}, \mathbf{W}_{5}$, and  $\mathbf{W}_{6}$ are trainable convolutional kernels. The resulting deep network has two distinct (model-based) non-linearities/activations per layer: the mixed $\ell_{1,2}$ thresholding, and singular value thresholding. The authors train the architecture end to end on a combination of simulations and RPCA results on real data, and demonstrate that it outperforms a strong non-model-based deep network (a ResNet). 

\subsubsection*{Deep unfolding for ultrasound localisation microscopy}
\label{sec:deep_unfolded_ulm}

In the spirit of unfolding, \citet{van2019deep} propose to unfold their sparse recovery algorithm for ULM to enable accurate localization even for high concentrations of microbubbles. Similar to the previous examples of unfolding, each of the layers $k$ in the resulting architecture takes the following form:
\begin{align}
    \label{eqn:ISTA_ULM}
    {\mathbf{x}}^{(k+1)}=&\mathcal{T}_{\lambda^{(k)}}\left(\mathbf{W}_1^{(k)}\mathbf{y} + \mathbf{W}_2^{(k)}\mathbf{x}^{(k)}\right),
\end{align}
with $\mathbf{W}_1^{(k)}$, $\mathbf{W}_2^{(k)}$ being trainable convolutional kernels. The authors train this convolutional LISTA architecture on simulated envelope-detected RF US data comprising microbubbles under a distribution of point-spread functions. This method was later adopted by \citet{bar2021learned}, who use it to perform localization microscopy in \textit{in-vivo} breast lesions. 

\citet{youn2021model} take this one step further, and combine the image-domain LISTA architecture by Van Sloun \textit{et al.} with the ABLE beamforming architecture by \citet{luijten2020adaptive}, training the joint network end-to-end. The authors show that this outperforms non-joint optimization, and that ABLE learns to accommodate the downstream localization problem addressed by LISTA. 

\subsection*{Speckle-suppression using deep generative priors}
\label{sec:DGM_speckle}

\noindent \cite{van2021ultrasound} formulate the task of speckle suppression as a MAP problem in which a clean image $\mathbf{x}$ is recovered from a measured, speckle-corrupted, image $\mathbf{y}$:
\begin{align*}
   \hat{\mathbf{x}} &= \arg\max_{\mathbf{x}} p(\mathbf{x}|\mathbf{y}) = \arg\max_{\mathbf{x}}  p(\mathbf{y}|\mathbf{x}) p(\mathbf{x}).
\end{align*}
The authors propose to model the clean image prior $p(\mathbf{x})$ using a deep generative model (a normalizing flow) trained on MRI images, which naturally have no speckle but display similar anatomical structure. Under such a deep generative normalizing flow prior (with normalized hidden space $\mathbf{z}$), optimization is then performed in $z$-space:
\begin{align*}
    \hat{\mathbf{z}} &= \arg\max_{\mathbf{z}} \log p(\mathbf{y}|f_\theta^{-1}(\mathbf{z}))+\log p(\mathbf{z}),
    \label{eqn:MAP1}
\end{align*}
where  $f^{-1}_\theta(\mathbf{z})$ is the inverse transformation of the normalizing flow $f_\theta(\mathbf{z})$, i.e. the generative direction. Assuming a simple Gaussian likelihood model for the log-compressed envelope detected US images, this can be rewritten as:
\begin{equation}
    \hat{\mathbf{z}} = \argmin_{\mathbf{z}} || f^{-1}_\theta(\mathbf{z}) - \mathbf{y}||_2^2 + \lambda||\mathbf{z}||_2^2,
    \label{eqn:MAPfinal}
\end{equation}
where $\lambda$ is a parameter that depends on the assumed noise variance. Iterative optimization of \eqref{eqn:MAPfinal} was performed using gradient descent, and the recovered clean image is given by $\hat{\mathbf{x}}=f^{-1}_\theta(\hat{\mathbf{z}})$.

\section*{Discussion}\label{sec:discussion}

Over the past decade, the field of ultrasound signal processing has seen a large transformation, with the development of novel algorithms and processing methods. This development is driven for a large part by the move from hardware- to software based reconstruction. In this review, we have showcased several works, from conventional algorithms, to full deep learning based approaches; each having their own strengths and weaknesses. 

Conventional model-based algorithms are based on first principles and offer a great amount of interpretability, which is relevant in clinical settings. However, as we show in this paper, these methods rely on estimations, and often simplifications of the underlying physics model, which result in sub-optimal signal reconstructions. For example, DAS beamforming assumes a linear measurement model, and a Gaussian noise profile, both of which are very crude approximations of a realistic ultrasound measurement. In contrast, adaptive methods (e.g. MV beamforming) that aim at modeling the signal statistics more accurately, are often computationally expensive to implement in real-time applications.

Spurred by the need to overcome these limitations, we see a shift in research towards data-driven signal processing methods (mostly based on deep learning), a trend that has started around 2014 \citep{zhang2021ai}, which sees a significant increase in the number of peer reviewed AI publications. This can be explained by 2 significant factors: 1) the availability of high compute-power GPUs, and 2) the availability of easy-to-use machine learning frameworks such as TensorFlow \citep{tensorflow2015-whitepaper} and PyTorch \citep{NEURIPS2019_9015}, which have significantly lowered the threshold of entry into the field of AI for ultrasound researchers.
However, the performance of data-driven, and more specifically, deep learning algorithms is inherently bounded by the availability of large amounts of high-quality training data. Acquiring ground truth data is not trivial in ultrasound beamforming and signal processing applications, and thus simulations or the outputs of advanced yet slow model-based algorithms are often considered as training targets.
Moreover, the lack of clear understanding of the behavior of learned models (i.e. the black box model), and ability to predict their performance ``in the wild'', makes implementations in clinical devices challenging.

These general challenges associated with fully data-driven deep learning methods have in turn spurred research in the field of ``model-based deep learning''. Model-based deep learning combines the model-based and data-driven paradigms, and offers a robust signal processing framework. It enables learning those aspects of full models from data for which no adequate first-principles derivation is available, or complementing/augmenting partial model knowledge.  Compared to conventional deep neural networks, these systems often require a smaller number of parameters, and less training data, in order to learn an accurate input-output mapping. 

Similar to model-based methods, we can broadly categorize model-based deep learning methods into: 1) algorithms based on iterative solutions, and 2) algorithms based on analytic solutions. Algorithms based on iterative solutions can be further split into: 1a) truncated algorithms (fixed number of iterations) of which (possibly a subset of) parameters are fine-tuned end-to-end (termed \textit{deep unfolding/unrolling}), and 1b) iterative algorithms with data-driven priors (e.g \textit{plug-and-play optimization}). Examples of 1a are deep unfolded ULM and deep unfolded robust PCA, but also unfolded ADMIRE beamforming. A recent example of 1b is the work by Van de Schaft \textit{et al.}, where MRI-based image priors learned with normalizing flows are used for ultrasound speckle suppression \citep{van2021ultrasound}. An example of 2) is ABLE, in which the analytic ML solution for beamforming under unknown non-diagonal covariance Gaussian channel noise is augmented with a neural network, and the entire hybrid solution is optimized end-to-end. The methods covered here aim to achieve a better imaging quality, e.g. temporal or spatial resolution, ultimately aiding in the diagnosis process. While a deeper analysis of the clinical relevance is a crucial and interesting topic, it is beyond the scope of this work.

\section*{Conclusion}
\label{sec:conclusion}

In this review, we outline the development of signal processing methods in US, from classic model-based algorithms, to fully data driven DL based methods. We also discuss methods that lie at the intersection of these two approaches, using neural architectures inspired by model-based algorithms, and derived from probabilistic inference problems. We take a probabilistic perspective, offering a generalised framework with which we can describe the multitude of approaches described in this paper, all under the same umbrella. This provides us insight into the demarcation between components derived from first principles, and the components derived from data. This also affords us the ability to combine these components in a unique combination, to derive architectures that integrate multiple classes of signal processing algorithms.

The application of such novel, DL based reconstruction methods requires the next generation of US devices to be equipped accordingly. Either by fast networking and on-device encoding, or by fully arming them with sufficient and appropriate processing power (GPUs and TPUs), which allows for flexible and real-time deployment of AI algorithms.

\section*{Acknowledgements}
\label{Ack}
This work was supported in part by the Dutch
Research Council (NWO) and Philips Research through the research
programme “High Tech Systems and Materials (HTSM)” under Project
17144.

\section*{Conflict of Interest}
\label{conflict}
The authors declare that there are no conflicts of interest in this work.





\pagebreak

\bibliographystyle{UMB-elsarticle-harvbib.bst}
\bibliography{bibliography}




\pagebreak

\end{document}